\documentclass[prx,aps,english,onecolumn,notitlepage,superscriptaddress]{revtex4-2}
\usepackage[T1]{fontenc}
\usepackage[latin1]{inputenc}
\usepackage{multirow,amsmath}
\usepackage{babel}
\usepackage{dsfont}
\usepackage[table]{xcolor}
\usepackage{txfonts}
\usepackage{epsfig,epsf,psfrag} \usepackage{adjustbox}
\usepackage{graphicx}
\usepackage{booktabs}
\usepackage{subfigure}
\usepackage{pslatex}
\usepackage{float}
\usepackage{upgreek}
\usepackage{subfigure}
\usepackage{epic,eepic}
\usepackage{color}
\usepackage{pstricks}
\usepackage{tabularx}
\usepackage{array}
\newcolumntype{Y}{>{\centering\arraybackslash}X}
\usepackage{physics}
\usepackage[colorlinks=true, allcolors=black]{hyperref}
\usepackage{braket}
\usepackage{url}
\usepackage{listings}
\usepackage{color}
\definecolor{lightgray}{gray}{0.85}
\definecolor{cream}{rgb}{1.0, 0.99, 0.82}
\definecolor{khaki(x11)(lightkhaki)}{rgb}{0.94, 0.9, 0.55}
\definecolor{vanilla}{rgb}{0.95, 0.9, 0.67}
\definecolor{ashgrey}{rgb}{0.7, 0.75, 0.71}
\definecolor{azure(web)(azuremist)}{rgb}{0.94, 1.0, 1.0}
\definecolor{maroon}{rgb}{0.5, 0.0, 0.0}
\begin{document}
\lstset{language=Mathematica}
\title{Resolving degeneracies in Google search via quantum stochastic walks}
\author{Colin Benjamin}\email{colin.nano@gmail.com}
\author{Naini Dudhe}\email{nainidudhe@gmail.com}
\affiliation{School of Physical Sciences, National Institute of Science Education and Research Bhubaneswar, Jatni 752050, India}
\affiliation{Homi Bhabha National Institute, Training School Complex, Anushaktinagar, Mumbai 400094, India}
\begin{abstract}
The internet is one of the most valuable technologies invented to date. Among them, Google is the most widely used search engine. The PageRank algorithm is the backbone of Google search, ranking web pages according to relevance and recency. We employ quantum stochastic walks (QSW) with the hope of bettering the classical PageRank (CPR) algorithm, which is based on classical continuous time random walks (CTRW). We implement QSW via two schemes: only incoherence and dephasing with incoherence. PageRank using QSW with only incoherence or QSW with dephasing and incoherence best resolves degeneracies that are unresolvable via CPR and with a convergence time comparable to that for CPR, which is generally the minimum. For some networks, the two QSW schemes obtain a convergence time lower than CPR and an almost degeneracy-free ranking compared to CPR.
\end{abstract}
\maketitle
\section{Introduction}
Google is the world's most popular search engine~\cite{whygooglebest,bestgoogle}. Most people do not give a second thought to what search engine to use. Whenever a question pops in the head to which the answer is unknown, the first thought is to `Google it. It sets the stage for the question: why is Google the most popular search engine? What makes it better than others? The question itself prompts the answer; Google must have an algorithm that gives better search results than other search engines~\cite{whygooglebest}. This algorithm is called the `PageRank' algorithm~\cite{rousseau2008mathematics,rogers2002google}. The PageRank algorithm ranks web pages according to the nature of the search and puts forth the most relevant web pages towards the top. The number of links to it determines the importance of a website. Of course, a good website will have more links to it from other pages.
Nevertheless, the number of links is not the only factor used for ranking. The quality of links matters as well, among other factors. Even if a website has fewer links than others, if the links are from well-known websites, it is likely to receive a higher rank than others~\cite{rogers2002google,pagerank}. The world wide web is a vastly complex network involving many websites and associated links. To rank, the web pages in such a vast network become pretty tricky, and the classical PageRank algorithm (CPR)~\cite{brin1998anatomy} might not be able to accurately distinguish between the importance of some websites, in turn giving them the same ranks. In other words, there might be degeneracies in the ranks of several websites. It becomes problematic as it lowers the search quality.

Several efforts have been made to improve the PageRank algorithm to produce better results~\cite{kim2002improved,langville2006reordering,mohan2017technique}. Since the classical PageRank algorithm has been mentioned, it provokes the thought: what if quantumness is introduced in the algorithm? Can it lead to some improvements? The authors in Ref.~\cite{sanchez2012quantum} tried to improve the CPR algorithm by using the approach of quantum stochastic walks (QSW). It was first put forth in Ref.~\cite{QSW_def} and can be said to be a hybrid of continuous-time quantum walks (CTQW) and classical continuous-time random walks (CTRW)~\cite{Portugal}. QSW uses a single parameter $\omega$ to interpolate between CTRW and CTQW, and its usefulness lies in the PageRank algorithm. We can estimate the value of $\omega$ for which we get the best results. Implementing the CPR algorithm involves administering CTRW on a network. It represents the probability of a user reaching a particular page by clicking on links randomly. The networks we have considered in this work are Erdos-Renyi, Watts-Strogatz, Scale-Free, and spatial networks, all possessing some properties of real-world networks. We use these networks as they are easy to study and can represent real-world networks to some degree. We use QSW in our quest to improve CPR instead of discrete-time quantum walks (DTQW)~\cite{coineddef} as graphs like Erdos-Renyi, Watts-Strogatz, Scale-Free, and spatial networks have a varying degree distribution of their verti  ces. Thus, implementing DTQW on such networks would imply a varying rank of the coin matrix operator. This makes implementing DTQW on such networks unwieldy, if not impossible. Further, QSW has the advantage that it can simulate a pure quantum evolution, i.e., CTQW~\cite{Farhi} as well as an utterly classical evolution, i.e., CTRW, for varying levels of quantumness or classicality.

This paper combines QSW with the PageRank algorithm to produce the quantum PageRank algorithm (QPR)~\cite{sanchez2012quantum,loke2017comparing} and rank webpages in different networks. Ref.~\cite{sanchez2012quantum} also studies different networks, such as ER and SF, using the QSW method of dephasing with incoherence. However, unlike us, their main focus is not on ranking sites in complex networks; instead, they focus on the convergence time. Furthermore, they only use dephasing with an incoherence scheme. At the same time, we rank sites for QSW methods of only incoherence and dephasing with incoherence and calculate convergence time for each case. We find that QSW methods with only incoherence or dephasing with incoherence have a degeneracy resolution significantly better than CPR, and a convergence time comparable to CPR, sometimes even better than CPR, making them a good choice for the PageRank algorithm. The paper's outline is as follows: in the next section, we introduce page ranking via CPR, which enables us to rank vertices in a network. After that, we introduce different QSW methods for producing the QPR algorithm. The results and Discussion section follow this, wherein we provide plots and tables which back up our claim that QPR with QSW methods of only incoherence and dephasing with incoherence is better suited to resolve degeneracies between vertex ranks as compared to CPR. We also calculate the convergence time here and find that for more extensive networks, the convergence time for QSW methods of only incoherence and dephasing with incoherence is comparable to that of CPR and sometimes even lesser than CPR. This makes either of them an appropriate choice for the PageRank algorithm. We then implement the CPR and QPR algorithms on smaller networks in the Analysis section to study the degeneracies between vertices at a smaller scale by analyzing ranks for each vertex which was impractical in more extensive networks due to them having a large number of vertices. We conclude our findings in the final section, where we provide a perspective on future endeavors. We end the paper with an Appendix that provides the \textit{Mathematica} codes used. 

\section{Theory}
\subsection{Classical continuous time random walks and classical PageRank}
Any complex network can be modeled as a graph of $M$ vertices with $E$ edges. For example, if we were to model a network of different websites, the vertices would represent the webpages, and the edges would represent the links from one webpage to the other~\cite{QSWalk}. In a random walk, a walker moves from one vertex $i$ to another vertex $j$ at each time step along an edge, if it exists, in a graph. In that case, the adjacency matrix will have the entry $A_{ij}=1$ and $A_{ji}=0$ if the edge is directed from $j$ to $i$ and if it is undirected, then $A_{ij}=A_{ji}=1$. The out-degree of a vertex is defined as the total number of edges leaving the vertex and is given by: $\text{out-deg}(j)=\sum_{i\neq j}A_{ij}$~\cite{QSWalk,Portugal}. The probability vector is a vector with dimension $M\times1$ and denotes the probability of the walker being at each vertex of the graph. The $i^{th}$ component of the probability vector (denoted as $p_i$ with $p_i \geq 0,\ \forall i$ and $\sum_i p_i=1$) represents the probability of the walker to be at vertex $i$ at time $t$. The probability for the walker to jump from vertex $i$ to $j$ is stored in the $(ji)^{th}$ entry of the transition matrix $\hat{T}$. The transition matrix has the property that all entries are non-negative real numbers, and the entries of any column must sum to $1$. For the discrete-time version where the time step is discrete, the probability vector evolves in time as,
\begin{equation}
\Vec{p}(t)=\hat{T}^t\Vec{p}(0),
\label{crweq}
\end{equation}
where $\Vec{p}(t)$ is the probability vector at time $t$ and $\hat{T}$ is the transition matrix. The transition matrix for the continuous time version is defined as (see Ref.~\cite{Portugal}),
\begin{equation}
T_{ij}=\begin{cases}
1-d_j\eta\epsilon+O(\epsilon^2), & i=j;\\
\eta\epsilon+O(\epsilon^2), & i\neq j.
\end{cases}
\label{transmat}
\end{equation}
Here, $d_j$ is the degree of the vertex $j$, $\eta$ is the probability of transition between neighboring vertices per unit time, and $\epsilon$ is the infinitesimal time step. Thus, the probability of the walker going from vertex $j$ to $i$ is $\eta\epsilon$. Now, in order to get a closed form expression for the evolution of the transition matrix in time, we define a generator matrix $\hat{H}$ with its $(ij)^{th}$ entry (see, Ref.~\cite{Portugal}) given by,
\begin{equation}
H_{ij}=\begin{cases}
d_j\gamma, & i=j;\\
-\gamma, & i\neq j \text{ and adjacent};\\
0, & i\neq j \text{ and non-adjacent}.
\end{cases}
\label{genmat}
\end{equation}
In the language of webpages, adjacent vertices or websites are those between whom an edge or a link exists. Now, multiplying the transition matrices at different times and rearranging the indices (see, Ref.~\cite{Portugal} for full derivation), we get,

\begin{equation}
\frac{dT_{ij}(t)}{dt}=-\sum_k H_{kj}T_{ik}(t).
\label{transevol}
\end{equation}
Solving Eq.~(\ref{transevol}) with the initial condition $\hat{T}(0)=\delta_{ij}$ (implying that initially, the user has an equal chance of being at any of the vertices), we get
\begin{equation}
\hat{T}(t)=e^{-\hat{H}t}.
\label{transevolform}
\end{equation}
Now, since $\hat{T}$ and $\hat{H}$ are symmetric, Eq.~(\ref{transevol}) reduces to,
\begin{equation}
\frac{dT_{ji}(t)}{dt}=-\sum_k H_{jk}T_{ki}(t).
\label{transevolsym}
\end{equation}
In order to find the probability vector at time $t$, we multiply $p_i(0)$ on both sides of Eq.~(\ref{transevolsym}) and sum over $i$, getting
\begin{equation}
\frac{ d(\sum_i (T_{ji}(t)p_i(0)))}{dt}=-\sum_k H_{jk}\sum_i (T_{ki}(t)p_i(0)),\mbox{ which gives } \frac{d p_j(t)}{dt}=-\sum_k H_{jk}p_k(t).
\label{crwconteq}
\end{equation}
Eq.~\ref{crwconteq} is the continuous-time version of the probability vector evolution. For CPR, the $i^{th}$ entry of the probability vector $\Vec{p}(t)$ signifies the probability of a user being at webpage $i$ at time $t$. Iterating Eq.~(\ref{crwconteq}) over and over again, we eventually reach the stationary state (see Ref.~\cite{sanchez2012quantum}),
\begin{equation}
\Vec{p}_{eq}=\lim_{t\rightarrow \infty}\Vec{p}(t),
\label{pstat}
\end{equation}
where $\Vec{p}_{eq}$ is the probability vector of the stationary state and $\Vec{p}(t)$ is the probability vector at time $t$. The steady state is defined as when the  norm of the difference between probability vectors at time $t=k$ and $t=k-1$, almost vanishes, i.e., $||\Vec{p}(t=k)-\Vec{p}(t=k-1)|| < \epsilon$, wherein $\epsilon$ is a very small number, we sort the resulting vector $\Vec{p}(t=k)=\Vec{p}_{eq}$ to give the page rank. The vertex (web page) with the highest rank will be the top search result, the second highest rank will be the second search result, and so on.

 The entries of this stationary solution represent the ranks of the different vertices. The rank of the vertex here represents the importance of a webpage. A higher-ranked webpage will come on top of the search.

On a graph $G$ with $M$ vertices and defined via the adjacency matrix $A$, the ``Google matrix" $\mathcal{G} $ for the graph is defined as see also Refs.~\cite{QSWalk,sanchez2012quantum},
\begin{equation}
\mathcal{G}_{ij}=\begin{cases}
\alpha A_{ij}/\text{out-deg}(j)+(1-\alpha)\frac{1}{M}, & \text{out-deg}(j)>0,\\
\frac{1}{M}, & \text{out-deg}(j)=0.
\end{cases}
\label{googlematrix}
\end{equation}
The factor $1/M$ is to avoid dead ends by introducing random jumps~\cite{QSWalk}. Continuously clicking on links from one website to another may lead us to a site with no further links to click, i.e., a dead end. In order to avoid that, random jumps are introduced, which take the user to another site chosen randomly. The factor $\alpha$ is the damping factor, and $0\leq\alpha\leq1$~\cite{QSWalk}. $\alpha$ controls the ratio of random jumps to the original edges of the graph. We want the random jumps to be minimal, and thus, a large value of $\alpha$ is needed. We take $\alpha$ to be 0.9 throughout this paper, similar to Ref.~\cite{sanchez2012quantum} (generally, it is taken to be 0.85). In the following subsection, we introduce the QSW methods used in the QPR algorithm.

\subsection{Quantum stochastic walks and quantum PageRank}
Quantum stochastic walk (QSW) is the hybrid version of CTRW and CTQW and was introduced in Ref.~\cite{QSW_def}. QSW can interpolate between these two walks using a parameter $\omega$. $\omega$ controls the ratio of coherent to incoherent dynamics and measures the environmental effect on quantum evolution. The Kossakowski-Lindblad master equation~\cite{KOSSAKOWSKI, Lindblad}, used to describe quantum stochastic processes and in modeling open quantum systems, is used to derive QSW QSW can be used to model quantum processes that are coupled to an environment. The master equation, which tells us how the walker evolves in time, is given by~\cite{QSW_def},
\begin{equation}
\frac{d\rho(t)}{dt}=-i\left (1-\omega \right)\left[H,{\rho(t)}\right]+\omega\sum_{x=1}^{X} \left(\hat{O}_{x}\rho(t)\hat{O}_{x}^{\dagger}-\frac{1}{2}\left(\hat{O}_{x}^{\dagger}\hat{O}_{x}\rho(t)+\rho(t)\hat{O}_{x}^{\dagger}\hat{O}_{x}\right)\right).
\label{QWS_def}
\end{equation}

Here, $\rho(t)$ is the density matrix representation of walker at time $t$ with dimension $M\times M$. $M$ is the total number of vertices with the corresponding basis states as $\{\ket{1}, . . . , \ket{M}\}$. Density matrix elements, $\rho_{ij}(t) = \bra{i} \rho(t) \ket{j}$. $H$ is the Hamiltonian operator which accounts for coherent system dynamics. The parameter $0\leq \omega\leq 1$ is responsible for interpolating between CTRW and CTQW. The operator $\hat{O}_x$ represents the Lindblad operator (sparse $M\times M$ matrices) with index $x$ going up to maximum value $X$. $X$ depends on the choice of the Lindblad operator, which represents scattering between a pair of vertices. The first term on the RHS of Eq.~(\ref{QWS_def}) indicates a contribution from coherent dynamics. In contrast, the second term accounts for a contribution from incoherent dynamics via the three methods: pure dephasing (PD), only incoherence (OI), and dephasing with incoherence (DI). A model for QSW with pure dephasing considers Lindblad operators involving only the diagonal elements of Google matrix $\mathcal{G}$, and therefore, the Lindblad operators are defined as (see section 4.2 of Ref.~\cite{QSWalk} and Refs.~\cite{kendon,QSW_def}),

\begin{equation}
\hat{O}_{x}={\sqrt{\mathcal{G}_{ii}}}\mbox{ } {\mid\!i\rangle\langle i\!\mid},
\label{pure-dep}
\end{equation}
where $x=1,2,...X$ and $X=M$ for pure dephasing case with $M$ being the total number of vertices in the Google matrix $\mathcal{G}$. Here, $\mathcal{G}_{ii}= \braket{i|\mathcal{G}|i}$ represents the diagonal elements of the Google matrix $\mathcal{G}$. By the term ``pure dephasing", we mean that classical hopping due to the second term in Eq.~(\ref{QWS_def}) only affects the diagonal elements of the Google matrix $\mathcal{G}$~\cite{sanchez2012quantum}. For $\omega=1$, Eq.~(\ref{QWS_def}) reduces to (see Ref.~\cite{QSWalk}),
\begin{equation}
\frac{d\rho_{ij}(t)}{dt}=\begin{cases}-\frac{\mathcal{G}_{ii}+\mathcal{G}_{jj}}{2}\rho_{ij}(t),& i \neq j, \\
0, & i=j.
\end{cases}
\label{purw=1}
\end{equation}\\
From Eq.~(\ref{purw=1}), it is apparent that with time, the off-diagonal elements of the density matrix die down exponentially while the diagonal elements, signifying the populations, remain constant. It implies that for the $\omega=1$ limit, the QSW method of pure dephasing does not reduce to CTRW, implying that the QSW method of pure dephasing cannot be used to rank vertices.

In QSW with only incoherence, the Lindblad operators involve only the off-diagonal elements of the Google matrix. The Lindblad operators are defined as, see Ref.~\cite{testingspeedup},
\begin{equation}
\begin{array}{cc}
\hat{O}_{x}={\sqrt{\mathcal{G}_{ij}}} \mbox{ } {\mid\!\!i\rangle\langle j\!\!\mid}, & i\neq j,
\end{array}
\label{incoh}
\end{equation}
where, $x=1,2,...X$ and $X=M(M-1)$ with $M$ being the total number of vertices in the Google matrix $\mathcal{G}$. Here, $\mathcal{G}_{ij}= \braket{i|\mathcal{G}|j}$ represents the off-diagonal elements of the Google matrix. Here, by the term ``only incoherence", we mean that classical hopping due to incoherent dynamics only affects the off-diagonal elements of the Google matrix $\mathcal{G}$~\cite{sanchez2012quantum}. In contrast to the pure dephasing scheme, for the only incoherent scattering scheme, in the $\omega=1$ limit, the QSW reduces to CTRW.\\
Finally, we model the QSW with dephasing and incoherence by including the Google matrix's diagonal and off-diagonal elements. Thus, the Lindblad operators reduce to ~\cite{QSW_def,QSWalk}:
\begin{equation}
\hat{O}_{x}={\sqrt{\mathcal{G}_{ij}}} \mbox{ } {\mid\!\!i\rangle\langle j\!\!\mid},
\label{dep-incoh}
\end{equation}
where $x=1,2,...X$ and $X=M^2$ for the dephasing with incoherence scheme since all elements of Google matrix are considered. Here, $\mathcal{G}_{ij}= \braket{i|\mathcal{G}|j}$ is the $(ij)^{th}$ element of the Google matrix. Here, by the term ``dephasing with incoherence", we mean that classical hopping affects all elements of the Google matrix $\mathcal{G}$~\cite{sanchez2012quantum}. For this scheme as well, in the $\omega=1$ limit, the QSW reduces to CTRW. In addition, in the $\omega=0$ limit, all three QSW schemes reduce to CTQW.

In order to solve for the time evolution of the density matrix $\rho$, the first step is to vectorize the density matrix as was done in Ref.~\cite{QSWalk}. The resulting vectorized equation is,
\begin{equation}
\frac{d\bar{\rho}}{dt}=\hat{\mathbf{O}}\cdot\bar{\rho}(t),
\label{rhovect}
\end{equation}
where, $\bar{\rho}(t)$ is the vectorized density matrix at time $t$ and the operator $\hat{\mathbf{O}}$ is given by (see Ref.~\cite{QSWalk}),
\begin{equation}
\hat{\mathbf{O}}=-i(1-\omega)(\mathds{1}_M\otimes H-H^T\otimes\mathds{1}_M)+\omega\sum_{x=1}^X\left({\hat{O}^*}_x\otimes\hat{O}_x-\frac{1}{2}\left(\mathds{1}_M\otimes{\hat{O}^\dag}_x\hat{O}_x+{\hat{O}^T}_x{\hat{O}^*}_x\otimes\mathds{1}_M\right)\right).
\label{vectoper}
\end{equation}
Here, $\omega$ has the same meaning as Eq.~(\ref{QWS_def}), $\mathds{1}_M$ is the $M\times M$ identity matrix with $M$ being the total number of vertices in the graph, $H$ is the Hamiltonian operator and $\hat{O}_x$ is the Lindblad operator as in Eq.~(\ref{QWS_def}). We get the closed form relation of the time evolution of the vectorized density matrix as (see Ref.~\cite{QSWalk}),
\begin{equation}
\bar{\rho}(t)=e^{\hat{\mathbb{O}}t}\cdot\bar{\rho}(0), \mbox{ where, $\bar{\rho}(0)$ is the initial density matrix.}
\label{rhorel}
\end{equation}

We use the QSW methods with either only incoherence or dephasing with incoherence to rank vertices in a network since the QSW scheme of pure dephasing, in its stationary state, returns the same rank for all vertices. Thus, the pure dephasing method is not a legitimate method for QPR. For all calculations related to page rank, we use different values for $\omega$, depending on which value gives us the shortest convergence time. By convergence time, we mean the time it takes the classical PageRank algorithm to converge to the equilibrium state $\Vec{p}_{eq}$ in Eq.~(\ref{pstat}) and, by analogy, to the equilibrium state $\rho_{eq}$ in quantum PageRank. Convergence time is explained in detail at the end of this section.
The same principle as done for classical page rank is followed in section II.B of our paper to calculate quantum page rank with the probability vector being replaced by the vectorized density matrix $\bar{\rho}(t)$ (see Eq.~\ref{rhovect} of the paper). We then evolve this vectorized density matrix in time via $\bar{\rho}(t)=e^{\hat{\mathbb{O}}t}\cdot\bar{\rho}(0), $ where, $\bar{\rho}(0)$ is the initial density matrix, with $\hat{\mathbf{O}}$ being the Lindblad operator. After evolving it for a certain time we reach the stationary or steady state  $\bar{\rho}_{eq}$ by which we mean  when the norm of the difference between the vectorized density matrix vectors at time $t$ and $t+1$, almost vanishes, i.e., $|| \bar{\rho}(t+1)- \bar{\rho}(t)|| < \epsilon$, with $\epsilon$ being a very small number. When the steady state is reached we sort the resulting vector $\bar{\rho}(t) = \bar{\rho}_{eq}$ which gives the quantum page rank.

We have studied ranking in several different networks, all classified under either Scale-Free (SF) or Watts-Strogatz (WS) or Erdos-Renyi (ER) or spatial networks. We analyze the quantum PageRank algorithm on these networks because they have properties of small-world networks, which is a proxy for internet networks on the world wide web. All of them have some properties of small-world networks. ER, WS, and SF networks differ based on the distribution $P(d)$ of the degree of the nodes $d$. ER networks have a Poisson distribution for the vertex degree, while the vertex degree in SF networks follows the distribution law: $P(d)\sim v^{-\eta}$, where $2<\eta<3$~\cite{sanchez2012quantum}. For Watts-Strogatz (WS) model~\cite{watts1998collective}, the vertex degree follows a modified Poisson distribution with an exponential tail~\cite{chen2018models}. We take the network examples of the following ER networks: Zachary Karate club~\cite{zachary1977information} and Bernoulli graph distribution~\cite{weisstein2002bernoulli}. Zachary Karate club is a social network of people belonging to a karate club from a university~\cite{zachary1977information}. Bernoulli graph distribution is a type of network where the degree distribution follows the Bernoulli distribution. The SF networks considered in this paper are: Barabasi Albert graph~\cite{albert2002statistical} and Price Graph distribution~\cite{price1965networks,price1976general}. The Barabasi Albert graph is generated based on an algorithm given by Barabasi and Albert to generate random SF networks. They designed the algorithm keeping in mind that the world wide web, internet, citation networks, etc., are approximately SF networks~\cite{albert2002statistical}. The Barabasi Albert graph can be considered a special case of the Price graph~\cite{price1965networks,price1976general}. Spatial networks can model neuron connectivity in the brain or short-range communication networks~\cite{barthelemy2011spatial}. Aside from these, we also consider some small networks with eight vertices, e.g., a random eight vertex graph~\cite{sanchez2012quantum} and randomly generated spatial and WS networks. We consider these smaller networks to study the degeneracies properly between vertices at a smaller scale. It is easier than applying it to larger graphs to study their degeneracies, such that comparing QPR methods with CPR is apparent.

Now, we move on to defining the degeneracies in a ranked network, i.e., a network with its vertices ranked via the PageRank algorithm. Two parameters define the efficiency of the PageRank algorithm, first, the degeneracy in ranks, and second, the convergence time. The better algorithm is the algorithm that returns the least degeneracies and converges faster.
\vspace{-0.2cm}
\subsubsection{Calculation of degeneracy}
\vspace{-0.2cm}
Internet query search devised by Google Page rank is a two-step process; first, based on your search query, a whole host of websites with matching content is listed in no particular order. Most pre-Google search engines, such as Lycos, Looksmart, etc., stopped here. The user looked through this list and found what was meaningful; s/he generally scrolled down sequentially. Google's masterstroke was in adding the second step, which was to rank these websites with the matching content based on links between them. To do this, it relied on a continuous time random walk. The website with the highest rank on this listed network of websites will be the website with a high number of links to itself, i.e., a highly connected vertex in a network, as well as one that possesses links from other highly ranked websites.

The PageRank (PR) algorithm (CPR or QPR) computes the ranking of each vertex in a graph and gives the output. Any vertex with more links to it will have a higher rank. Furthermore, any vertex having links from a higher ranked vertex will, in turn, have a high-rank itself~\cite{rogers2002google,pagerank}. Unavoidably, some of the vertices have the possibility of getting the same rank as they might have the same number of links from other websites, and the quality of the links might be similar too. It is undesirable since the user might have difficulties perceiving which website is more relevant. It gives rise to poor search quality. Thus, our aim in this work is to find a scheme that eliminates degeneracies between ranks of vertices. To this end, we count the total number of all these possible degeneracies using CPR as well as QPR with the two QSW schemes and compare them.
\vspace{-0.2cm}
\subsubsection{Convergence time}
\vspace{-0.2cm}
The time it takes the PageRank algorithm to converge to the stationary solution $\Vec{p}_{eq}$ given in Eq.~(\ref{pstat}) for CPR and $\rho_{eq}$ for QPR is called the convergence time. It is defined such that given any $\epsilon>0$, for $t>\tau$, we have $||\rho (t)-\rho_{eq}||<\epsilon$, where $||\rho||=\sqrt{\sum_i{(\rho_{ii}(t)-{\rho_{eq}}_{ii})}^2}$~\cite{sanchez2012quantum}. Since at $\omega=1$, the two QSW schemes reduce to CPR, we consider the convergence time for the QSW scheme of dephasing with incoherence at $\omega=1$ to be the convergence time of CPR~\cite{sanchez2012quantum}. The convergence time has the upper bound ${|Re[\lambda_1]|}^{-1}$, where $\lambda_1$ is the largest eigenvalue of the operator $\hat{\mathbb{O}}$ in Eq.~(\ref{vectoper}) (see Ref.~\cite{sanchez2012quantum}). If convergence time for QPR is better than CPR, then we expect $\tau_{QPR}<\tau_{CPR}$. We plot the ratio of convergence time for QPR with that for CPR, i.e., QSW at $\omega=1$ limit ($\tau_{QPR}/\tau_{CPR}$).
\vspace{-0.2cm}
\section{Results and Discussion}
\vspace{-0.2cm}
We have studied four types of networks in this work: Erdos-Renyi (ER) networks, Watts-Strogatz (WS) model, Scale-Free (SF) networks, and spatial networks. These networks have around 100 vertices each. The similarity between these is that they all have some properties of small world networks~\cite{amaral2000classes}. Small world networks possess three essential properties: (i) any two nodes will have a relatively small distance between them (distance here means that in order to get to one node from the other, the walker will not have to pass through too many nodes), (ii) clustering is high and (iii) hubs are present. Here, hubs are the vertices possessing a large degree. In turn, the degree distribution will have a long tail since the hubs will have a much higher degree than others. Also, in order to explain clustering, let us assume one node $A$ is connected to two other nodes $B$ and $C$; in that case, the chances for $B$ and $C$ to be joined via an edge are high, i.e., $A$, $B$ and $C$ are mutually connected. It is called clustering. Some examples of small-world networks are road maps, electric power grids, social networks, neural networks, and even the internet~\cite{watts1998collective}. ER, SF and WS networks all have the property that any two nodes will have relatively less distance between them. However, ER networks do not have clustering or the presence of hubs (the degree distribution is Poissonian and does not have a long tail).

On the other hand, SF networks possess hubs, but clustering is not that high in these. The WS model proposed by Watts and Strogatz possesses a high clustering coefficient. However, the degree distribution is unrealistic, not similar to real networks~\cite{chehreghani2014modeling} due to a Poisson distribution with an exponential tail and not a long tail like SF networks~\cite{chen2018models}. There also exist some types of networks where the distance between nodes greatly influences the probability of them being connected, and thus, they can be aptly explained using spatial networks. Like communication networks having short radio ranges, people having friends and relatives in their neighborhoods can be modeled using spatial networks. Connectivity in the brain depends on the distance between neurons. Closer regions have a higher probability of being connected than far apart regions. These are all examples of spatial networks~\cite{barthelemy2011spatial}.

The reason for studying these four types of networks is because each can be used to model some real network, and they are relatively easy to study. To this end, we rank vertices in these networks and check for the number of degeneracies to find the most feasible search algorithm that can eliminate these degeneracies. In addition, we compute the convergence time to narrow it down to an algorithm capable of getting rid of degeneracies in the least time. We consider two examples of ER networks: the Zachary Karate club and Bernoulli graph distribution; the WS network is sui-generis, while we study two examples for SF networks, the Barabasi-Albert distribution, and Price graph distribution. Similarly, we analyze two spatial network examples differing in distance measures. Distance measure in spatial networks is the maximum distance between two nodes for an edge to exist between them. For example, in communication networks, there is a range, crossing, which results in disturbance in the signal. Thus, if the distance measure is small, the network has a short range and will only work correctly when vertices are near each other. In other words, vertices will have edges connecting them if they are near each other (a sparsely connected network). On the other hand, if the distance measure is high, the network has a high range and can work correctly even if vertices are far apart. This means that the network will be highly connected, as edges can exist between vertices even if they are far apart.
\vspace{-0.3cm}
\subsection{Erdos-Renyi networks}
\vspace{-0.8cm}
\begin{figure}[H]
\centering
\includegraphics[width=\textwidth]{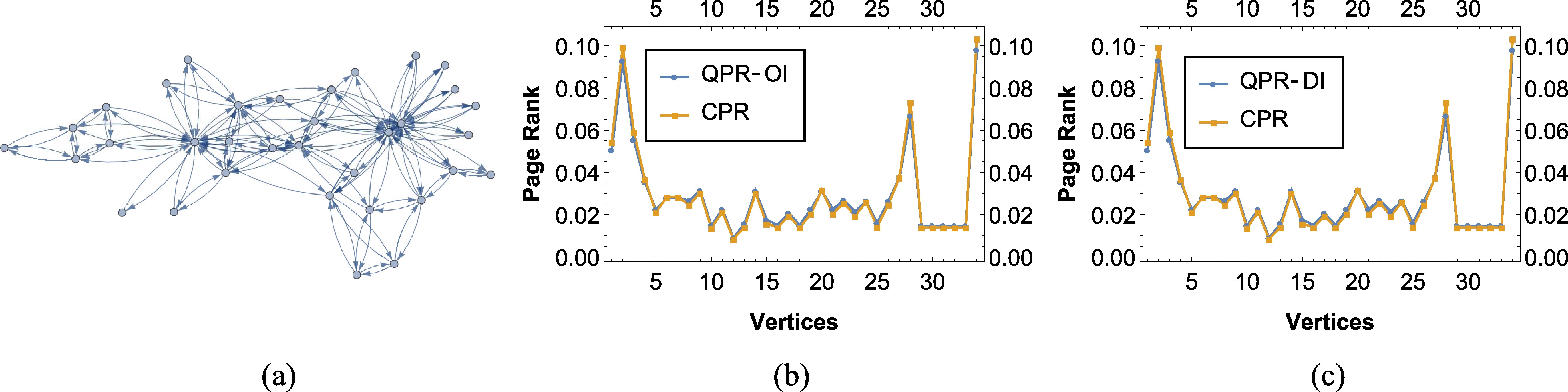}
\vspace{-0.6cm}
\caption{(a) The ER network Zachary Karate club. This network is inbuilt in \textit{Mathematica}, see Ref.~\cite{zacharynetwork}. Ranks for each vertex for QSW with (b) only incoherence (QPR-OI) and (c) dephasing with incoherence (QPR-DI) for ER network Zachary Karate club. CPR values are also given for comparison.}
\label{ER_zachary}
\end{figure}
\vspace{-0.6cm}
\begin{figure}[H]
\centering
\includegraphics[width=\textwidth]{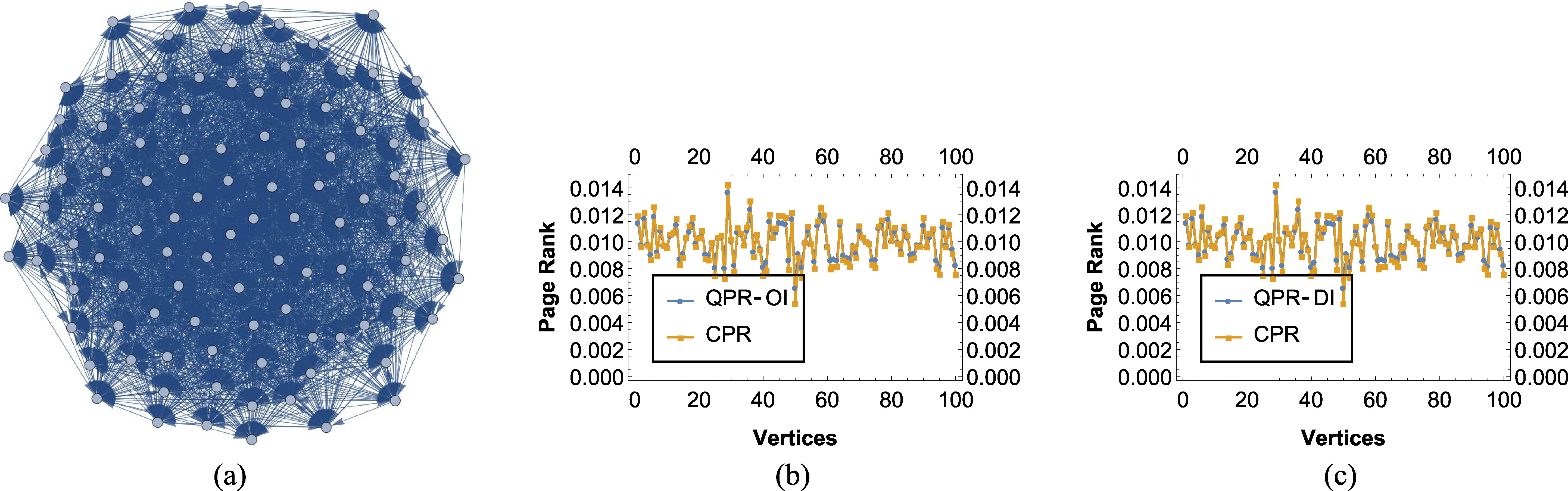}
\vspace{-0.6cm}
\caption{(a) A randomly generated ER network Bernoulli graph distribution of 100 vertices. Bernoulli graph distribution is available in \textit{Mathematica} (see Ref.~\cite{bernoulligraph}). BernoulliGraphDistribution[n,p] in \textit{Mathematica} shows a graph of $n$ vertices and the probability of an edge existing $p$. We have used BernoulliGraphDistribution[100,0.6]. Ranks for each vertex for QSW with (b) only incoherence (QPR-OI) and (c) dephasing with incoherence (QPR-DI) for a randomly generated Bernoulli graph distribution. CPR values are also given for comparison.}
\label{ER_bernoulli}
\end{figure}
\vspace{-0.5cm}
\begin{table}[H]
\centering
\begin{tabular}{|ccc|ccc|}
\hline\hline
\multicolumn{3}{|c|}{\multirow{4}{*}{ER networks}} & \multicolumn{3}{c|}{Number of degeneracies} \\ \cline{4-6}
\multicolumn{3}{|c|}{} & \multicolumn{1}{c|}{\multirow{3}{*}{CPR}} & \multicolumn{2}{c|}{QPR} \\ \cline{5-6}
\multicolumn{3}{|c|}{} & \multicolumn{1}{c|}{} & \multicolumn{1}{c|}{\begin{tabular}[c]{@{}c@{}}Only\\ incoherence\end{tabular}} & \begin{tabular}[c]{@{}c@{}}Dephasing with\\ incoherence\end{tabular} \\ \hline
\multicolumn{1}{|c|}{34 vertices} &
\multicolumn{2}{c|}{\begin{tabular}[c]{@{}c@{}}Zachary\\ Karate club\\ network\end{tabular}} & \multicolumn{1}{c|}{7} & \multicolumn{1}{c|}{7} & 7 \\ \hline
\multicolumn{1}{|c|}{\multirow{5}{*}{\begin{tabular}[c]{@{}c@{}}100 vertices\end{tabular}}} & \multicolumn{1}{c}{\multirow{5}{*}{\begin{tabular}[c]{@{}c@{}}Bernoulli\\ graph\\ distribution\end{tabular}}} & \multicolumn{1}{|c|}{Network 1} & \multicolumn{1}{c|}{0} & \multicolumn{1}{c|}{0} & 0 \\ \cline{3-6}
\multicolumn{1}{|c|}{} & & \multicolumn{1}{|c|}{Network 2} & \multicolumn{1}{c|}{0} & \multicolumn{1}{c|}{0} & 0 \\ \cline{3-6}
\multicolumn{1}{|c|}{} & & \multicolumn{1}{|c|}{Network 3} & \multicolumn{1}{c|}{0} & \multicolumn{1}{c|}{0} & 0 \\ \cline{3-6}
\multicolumn{1}{|c|}{} & & \multicolumn{1}{|c|}{Network 4} & \multicolumn{1}{c|}{0} & \multicolumn{1}{c|}{0} & 0 \\ \cline{3-6}
\multicolumn{1}{|c|}{} & & \multicolumn{1}{|c|}{Network 5} & \multicolumn{1}{c|}{0} & \multicolumn{1}{c|}{0} & 0 \\ \hline\hline
\end{tabular}
\caption{comparison of the number of degeneracies in different ER networks- CPR versus QPR (using both QSW schemes). Five random graphs are obtained for the Bernoulli graph distribution. Both Q.P.R. schemes have identical degeneracy count, which is the same as CPR.}
\label{ERcomp}
\end{table}
\begin{figure}[H]
\centering
\includegraphics[width=0.7\textwidth]{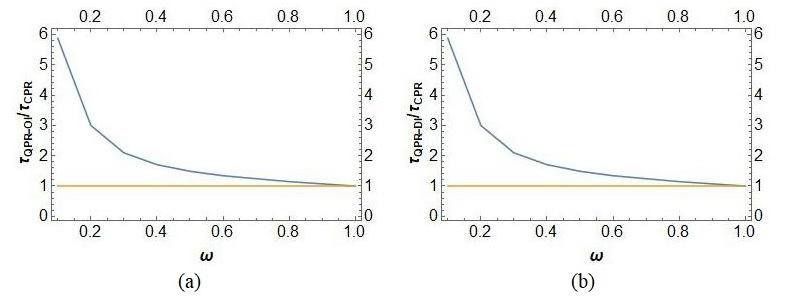}
\vspace{-0.4cm}
\caption{Convergence time $\tau_{QPR}/\tau_{CPR}$ versus $\omega$ for QSW with (a) only incoherence ($\tau_{QPR-OI}/\tau_{CPR}$) and (b) dephasing with incoherence ($\tau_{QPR-DI}/\tau_{CPR}$) for the ER network Zachary Karate club. Convergence times for CPR are better than QPR.}
\label{tau_zachary}
\end{figure}
\begin{figure}[H]
\centering
\includegraphics[width=0.7\textwidth]{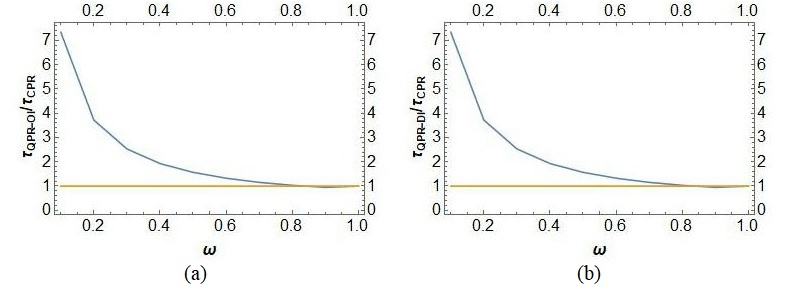}
\vspace{-0.4cm}
\caption{Convergence time $\tau_{QPR}/\tau_{CPR}$ versus $\omega$ for QSW with (a) only incoherence ($\tau_{QPR-OI}/\tau_{CPR}$) and (b) dephasing with incoherence ($\tau_{QPR-DI}/\tau_{CPR}$) for the ER network Bernoulli graph distribution. The plots have been averaged over 5 randomly generated Bernoulli graph distribution with $n=100$, $p=0.6$, same as in Fig.~\ref{ER_bernoulli}. Convergence time for QPR is marginally better than CPR at $\omega=0.9$.}
\label{tau_bernoulli}
\end{figure}
Fig.~\ref{ER_zachary}(a) depicts the ER network Zachary Karate club, which consists of 34 vertices. Similarly, Fig.~\ref{ER_bernoulli}(a) depicts the ER network Bernoulli graph distribution (randomly generated) of 100 vertices. Fig.~\ref{ER_zachary}(b-c) shows the ranks of all vertices using CPR and QPR (for both QSW schemes) pictorially for the Zachary Karate club network, while Fig.~\ref{ER_bernoulli}(b-c) depicts the same for Bernoulli graph distribution. In Table~\ref{ERcomp}, we compute the number of degeneracies in the Zachary Karate club network and Bernoulli graph distribution. From Table~\ref{ERcomp}, for the Zachary Karate club network, one can see that using CPR and QSW with either scheme leads to seven degeneracies. On the other hand, for Bernoulli graph distribution, CPR shows no degeneracies. QPR also shows no degeneracies for this case. The plots for $\tau_{QPR}/\tau_{CPR}$ versus $\omega$ for both the QSW schemes are given in Fig.~\ref{tau_zachary} for the ER network Zachary Karate club and in Fig.~\ref{tau_bernoulli} for the ER network Bernoulli graph distribution. For Zachary Karate club, the convergence time curve is decreasing, achieving a minimum value at $\omega=1$, i.e., CPR. For Bernoulli graph distribution, the convergence time curve $\tau_{QPR}$ decreases up to $\omega=0.9$, and then increases slightly at $\omega=1$ to $\tau_{CPR}$. For both ER networks, while checking for degeneracies, we consider $\omega$ to be 0.9. It makes sense when we take this value of $\omega$ for the Bernoulli graph distribution. Nevertheless, our reason for taking $\omega=0.9$ for Zachary Karate club is that we wish to see the effect of QPR, which cannot be seen if we assume $\omega=1$ as it reduces to CPR. Still, taking this $\omega$ value is all right since the difference in convergence time at $\omega=0.9$ or $1$ is negligible. There is no difference when comparing degeneracies between CPR and QSW via any of the two schemes. Thus, there is little to choose between QPR and CPR as regards resolving degeneracies for ER networks. Ref.~\cite{sanchez2012quantum} also analyzes ER networks, but unlike us, they only focus on convergence time. They take an average of over 50 ER networks with 200 nodes and find an optimal convergence time $\tau_{QR}$ which is better than $\tau_{CPR}$ while we compute both the convergence time as well as the degeneracy count. Additionally, they do not specify what ER networks they use while we do.
\subsection{Watts-Strogatz network}
\vspace{-0.5cm}
\begin{figure}[H]
\centering
\includegraphics[width=\textwidth]{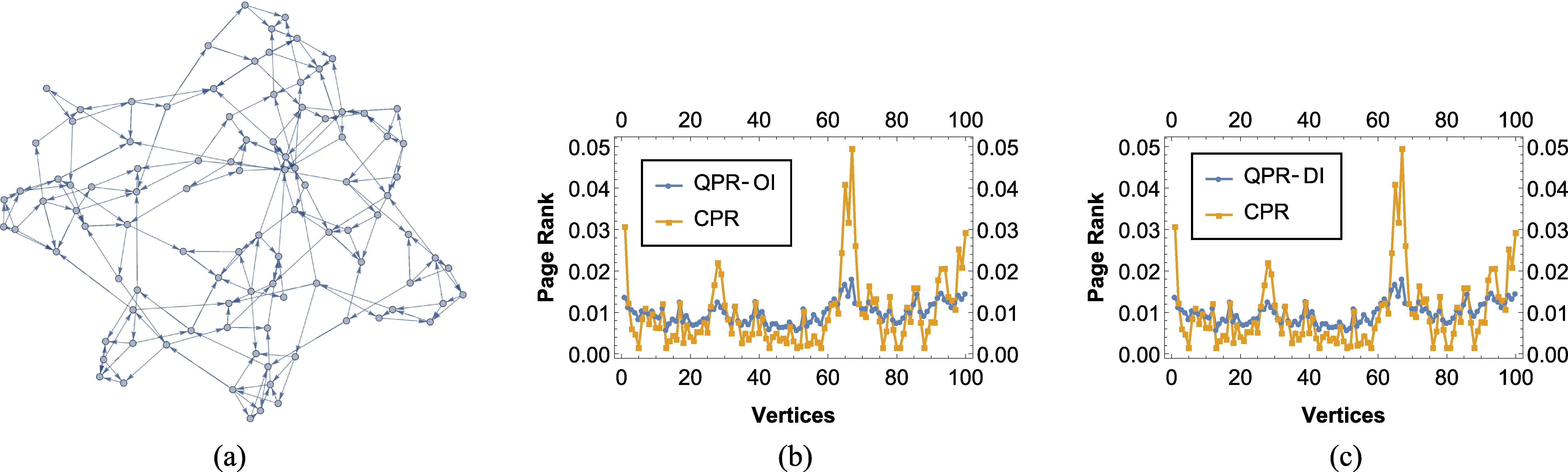}
\caption{(a) A randomly generated WS network of 100 vertices. WattsStrogatzGraphDistribution[n,p] in \textit{Mathematica} shows a graph of $n$ vertices and the probability of rewiring $p$. We have used WattsStrogatzGraphDistribution[100,0.2]. Ranks for each vertex for QSW with (b) only incoherence (QPR-OI) and (c) dephasing with incoherence (QPR-DI) for a randomly generated WS network. CPR values are also given for comparison.}
\label{ER_watts}
\end{figure}
\vspace{-0.5cm}
\begin{table}[H]
\centering
\begin{tabular}{|cc|ccc|}
\hline\hline
\multicolumn{2}{|c|}{\multirow{4}{*}{Watts-Strogatz (WS) network}} & \multicolumn{3}{c|}{Number of degeneracies} \\ \cline{3-5}
\multicolumn{2}{|c|}{} & \multicolumn{1}{c|}{\multirow{3}{*}{CPR}} & \multicolumn{2}{c|}{QPR} \\ \cline{4-5}
\multicolumn{2}{|c|}{} & \multicolumn{1}{c|}{} & \multicolumn{1}{c|}{\begin{tabular}[c]{@{}c@{}}QSW (Only\\ incoherence)\end{tabular}} & \begin{tabular}[c]{@{}c@{}}QSW (Dephasing\\with incoherence)\end{tabular} \\ \hline
\multicolumn{1}{|c|}{\multirow{5}{*}{\begin{tabular}[c]{@{}c@{}}100 vertices\end{tabular}}} & Network 1 & \multicolumn{1}{c|}{6} & \multicolumn{1}{c|}{0} & 1 \\ \cline{2-5}
\multicolumn{1}{|c|}{} & Network 2 & \multicolumn{1}{c|}{8} & \multicolumn{1}{c|}{2} & 0 \\ \cline{2-5}
\multicolumn{1}{|c|}{} & Network 3 & \multicolumn{1}{c|}{10} & \multicolumn{1}{c|}{4} & 3 \\ \cline{2-5}
\multicolumn{1}{|c|}{} & Network 4 & \multicolumn{1}{c|}{8} & \multicolumn{1}{c|}{2} & 1 \\ \cline{2-5}
\multicolumn{1}{|c|}{} & Network 5 & \multicolumn{1}{c|}{9} & \multicolumn{1}{c|}{3} & 2 \\ \hline\hline
\end{tabular}
\caption{comparison of the number of degeneracies in WS network- CPR versus QPR (using both QSW schemes). Five random graphs are obtained. Both Q.P.R. schemes show much less degeneracy count.}
\label{WScomp}
\end{table}
\vspace{-0.2cm}
\begin{figure}[H]
\centering
\includegraphics[width=0.7\textwidth]{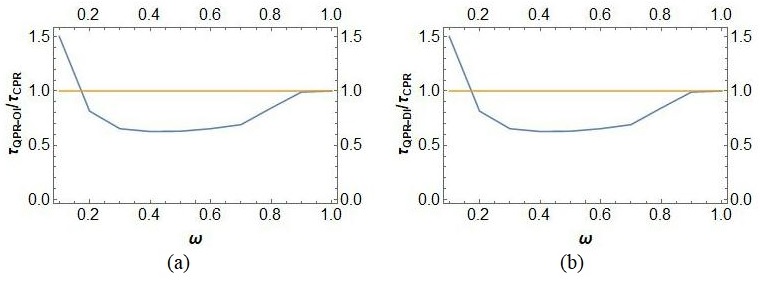}
\vspace{-0.5cm}
\caption{Convergence time $\tau_{QPR}/\tau_{CPR}$ versus $\omega$ for QSW with (a) only incoherence ($\tau_{QPR-OI}/\tau_{CPR}$) and (b) dephasing with incoherence ($\tau_{QPR-DI}/\tau_{CPR}$) for the WS network. The plots have been averaged over five randomly generated WS networks with $n,p$ as in Fig.~\ref{ER_watts}. Both Q.P.R. schemes at $\omega=0.4$ have convergence times much less than CPR.}
\label{tau_WS}
\end{figure}
A randomly generated WS network of 100 vertices is shown in Fig.~\ref{ER_watts}(a). To construct a WS graph, one initially starts with a ring lattice and then rewires some edges, which means that some edges are modified. Thus the first node remains the same while the second node is chosen randomly~\cite{song2014simple}. Fig.~\ref{ER_watts}(b-c) depicts the ranks of all vertices pictorially for both QSW schemes and CPR. In Table~\ref{WScomp}, we compute the number of degeneracies in the WS network. It shows an average of around eight degeneracies using CPR, while QPR using the QSW method of only incoherence gives an average of around two degeneracies, and the QSW method of dephasing with incoherence gives a degeneracy between 1 and 2. The plots for $\tau_{QPR}/\tau_{CPR}$ versus $\omega$ for both QSW schemes are given in Fig.~\ref{tau_WS}. The convergence time decreases up to $\omega=0.4$ and then increases. Thus, while checking for degeneracies for the two QSW methods, we take $\omega=0.4$ since the convergence time is minimum for this value of $\omega$. Thus, $\tau_{QPR}$ is much less than $\tau_{CPR}$, meaning quantum advantage is apparent in both convergence time and degeneracy removal via the QPR algorithm in WS networks.
\subsection{Scale-Free networks}
\vspace{-0.5cm}
\begin{figure}[H]
\centering
\includegraphics[width=\textwidth]{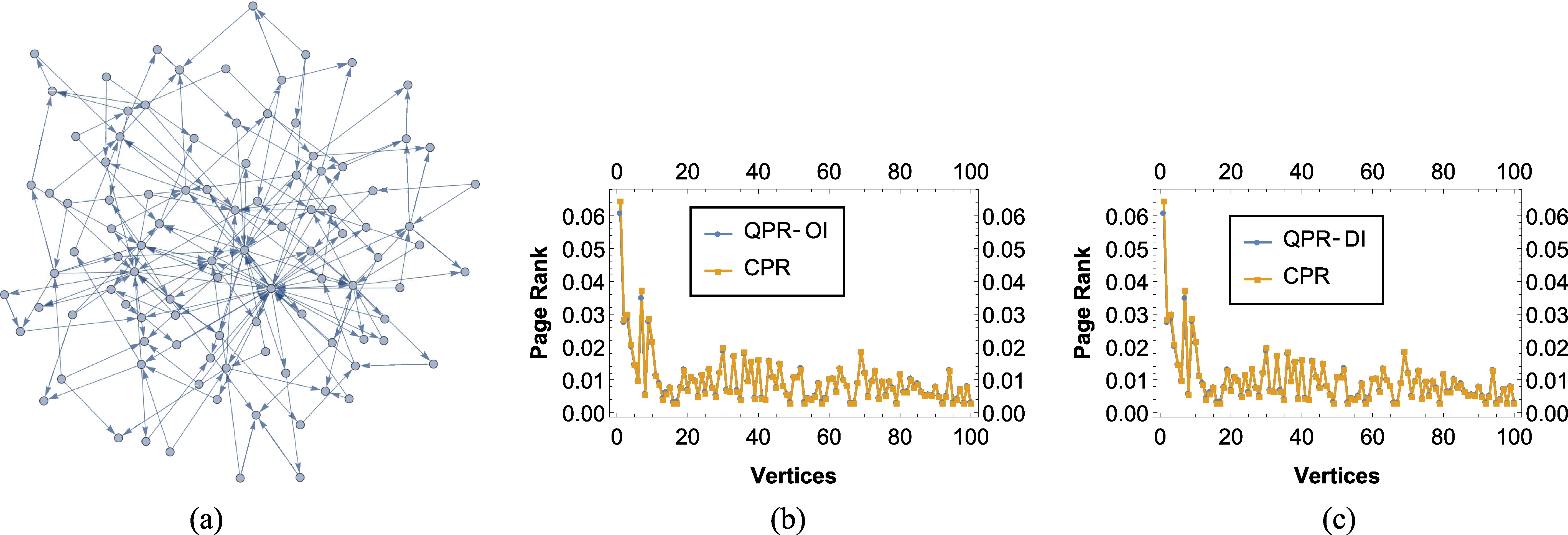}
\vspace{-0.3cm}
\caption{(a) A randomly generated SF network Barabasi-Albert graph distribution. BarabasiAlbertGraphDistribution[n,k] in \textit{Mathematica} shows a graph of $n$ vertices. We start with a single vertex, and a new vertex with $k$ edges is added at each step. We have used BarabasiAlbertGraphDistribution[100,2]. Ranks for each vertex for QSW with (b) only incoherence (QPR-OI) and (c) dephasing with incoherence (QPR-DI) for a randomly generated SF network with Barabasi-Albert distribution. CPR values are also given for comparison.}
\label{SF_barabasi}
\end{figure}
\vspace{-0.8cm}
\begin{figure}[H]
\centering
\includegraphics[width=\textwidth]{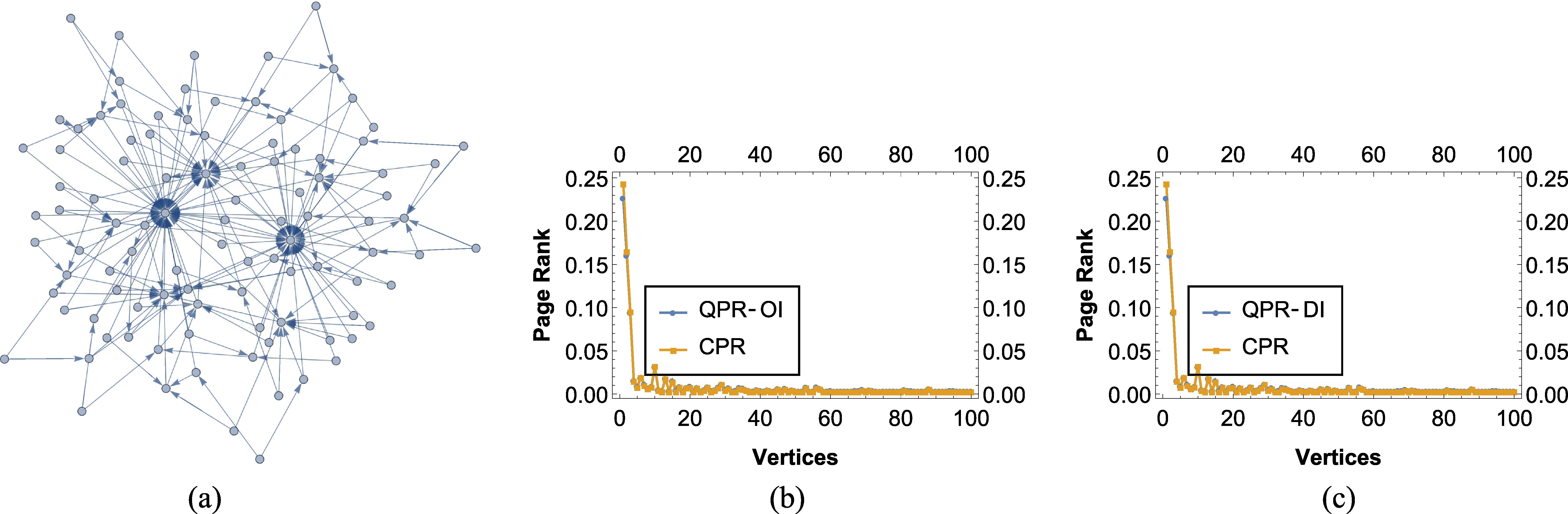}
\vspace{-0.3cm}
\caption{(a) A randomly generated SF network Price graph distribution. PriceGraphDistribution[n,k,a] in \textit{Mathematica} shows a graph of $n$ vertices. Here, we start with a single vertex, and a new vertex with $k$ edges is added at each step with weights $q_i+a$. Here, $q_i$ is the in-degree of vertex $i$. We have used PriceGraphDistribution[100,2,1]. Ranks for each vertex for QSW with (b) only incoherence (QPR-OI) and (c) dephasing with incoherence (QPR-DI) for a randomly generated SF network with Price graph distribution. CPR values are also given for comparison.}
\label{SF_price}
\end{figure}
\vspace{-0.8cm}
\begin{table}[H]
\centering
\begin{tabular}{|cc|ccc|}
\hline\hline
\multicolumn{2}{|c|}{\multirow{4}{*}{SF networks}} & \multicolumn{3}{c|}{Number of degeneracies} \\ \cline{3-5}
& & \multicolumn{1}{c|}{\multirow{3}{*}{CPR}} & \multicolumn{2}{c|}{QPR} \\ \cline{4-5}
& & \multicolumn{1}{c|}{} & \multicolumn{1}{c|}{\begin{tabular}[c]{@{}c@{}}QSW (Only\\ incoherence)\end{tabular}} & \multicolumn{1}{c|}{\begin{tabular}[c]{@{}c@{}}QSW (Dephasing\\ with incoherence)\end{tabular}} \\ \hline
\multicolumn{1}{|c|}{\multirow{10}{*}{\begin{tabular}[c]{@{}c@{}}Barabasi\\ Albert\\ graph\\ distribution\end{tabular}}} & Network 1 & \multicolumn{1}{c|}{33} & \multicolumn{1}{c|}{1} & 0 \\ \cline{2-5}
\multicolumn{1}{|c|}{} & Network 2 & \multicolumn{1}{c|}{32} & \multicolumn{1}{c|}{1} & 1 \\ \cline{2-5}
\multicolumn{1}{|c|}{} & Network 3 & \multicolumn{1}{c|}{25} & \multicolumn{1}{c|}{1} & 2 \\ \cline{2-5}
\multicolumn{1}{|c|}{} & Network 4 & \multicolumn{1}{c|}{30} & \multicolumn{1}{c|}{1} & 1 \\ \cline{2-5}
\multicolumn{1}{|c|}{} & Network 5 & \multicolumn{1}{c|}{29} & \multicolumn{1}{c|}{3} & 2 \\ \cline{2-5}
\multicolumn{1}{|c|}{} & Network 6 & \multicolumn{1}{c|}{30} & \multicolumn{1}{c|}{2} & 1 \\ \cline{2-5}
\multicolumn{1}{|c|}{} & Network 7 & \multicolumn{1}{c|}{35} & \multicolumn{1}{c|}{3} & 4 \\ \cline{2-5}
\multicolumn{1}{|c|}{} & Network 8 & \multicolumn{1}{c|}{31} & \multicolumn{1}{c|}{3} & 3 \\ \cline{2-5}
\multicolumn{1}{|c|}{} & Network 9 & \multicolumn{1}{c|}{28} & \multicolumn{1}{c|}{0} & 0 \\ \cline{2-5}
\multicolumn{1}{|c|}{} & Network 10 & \multicolumn{1}{c|}{28} & \multicolumn{1}{c|}{0} & 0 \\ \hline
\multicolumn{1}{|c|}{\multirow{10}{*}{\begin{tabular}[c]{@{}c@{}}Price\\ graph\\ distribution\end{tabular}}} & Network 1 & \multicolumn{1}{c|}{72} & \multicolumn{1}{c|}{5} & 5 \\ \cline{2-5}
\multicolumn{1}{|c|}{} & Network 2 & \multicolumn{1}{c|}{77} & \multicolumn{1}{c|}{11} & 10 \\ \cline{2-5}
\multicolumn{1}{|c|}{} & Network 3 & \multicolumn{1}{c|}{74} & \multicolumn{1}{c|}{5} & 5 \\ \cline{2-5}
\multicolumn{1}{|c|}{} & Network 4 & \multicolumn{1}{c|}{78} & \multicolumn{1}{c|}{13} & 13 \\ \cline{2-5}
\multicolumn{1}{|c|}{} & Network 5 & \multicolumn{1}{c|}{77} & \multicolumn{1}{c|}{15} & 13 \\ \cline{2-5}
\multicolumn{1}{|c|}{} & Network 6 & \multicolumn{1}{c|}{78} & \multicolumn{1}{c|}{14} & 14 \\ \cline{2-5}
\multicolumn{1}{|c|}{} & Network 7 & \multicolumn{1}{c|}{72} & \multicolumn{1}{c|}{8} & 7 \\ \cline{2-5}
\multicolumn{1}{|c|}{} & Network 8 & \multicolumn{1}{c|}{74} & \multicolumn{1}{c|}{6} & 7 \\ \cline{2-5}
\multicolumn{1}{|c|}{} & Network 9 & \multicolumn{1}{c|}{78} & \multicolumn{1}{c|}{14} & 15 \\ \cline{2-5}
\multicolumn{1}{|c|}{} & Network 10 & \multicolumn{1}{c|}{75} & \multicolumn{1}{c|}{8} & 12 \\ \hline\hline
\end{tabular}
\caption{comparison of the number of degeneracies in different SF networks, with 100 vertices each- CPR versus QPR (using both QSW schemes). Ten random distributions are obtained for both SF networks. Both QPR schemes show much less degeneracy count as compared to CPR.}
\label{SFcomp}
\end{table}
\begin{figure}[H]
\centering
\includegraphics[width=0.7\textwidth]{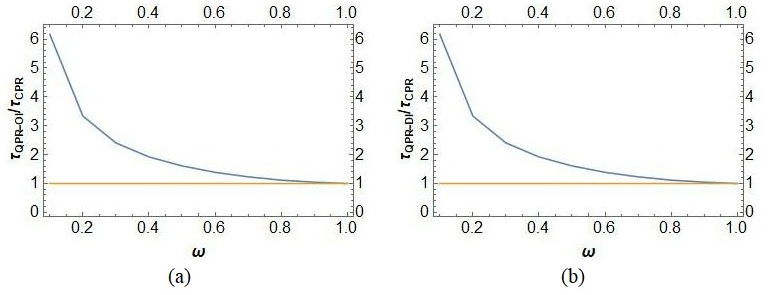}
\vspace{-0.4cm}
\caption{Convergence time $\tau_{QPR}/\tau_{CPR}$ versus $\omega$ for QSW with (a) only incoherence ($\tau_{QPR-OI}/\tau_{CPR}$) and (b) dephasing with incoherence ($\tau_{QPR-DI}/\tau_{CPR}$) for the SF network Barabasi-Albert graph distribution. The plots have been averaged over five randomly generated Barabasi-Albert distributions with the same parameters as in Fig.~\ref{SF_barabasi}. QPR convergence time at $\omega=0.9$ is marginally worse than CPR.}
\label{tau_BA}
\end{figure}
\vspace{-0.8cm}
\begin{figure}[H]
\centering
\includegraphics[width=0.7\textwidth]{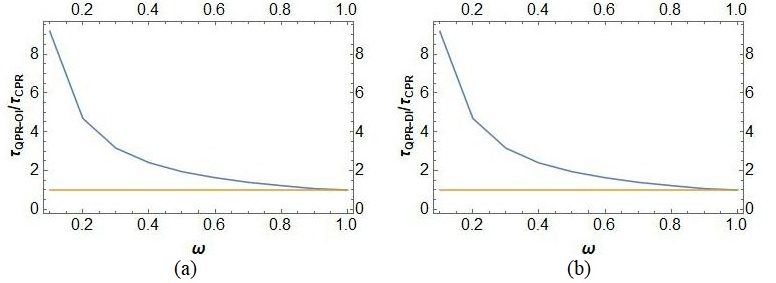}
\vspace{-0.4cm}
\caption{Convergence time $\tau_{QPR}/\tau_{CPR}$ versus $\omega$ for QSW with (a) only incoherence ($\tau_{QPR-OI}/\tau_{CPR}$) and (b) dephasing with incoherence ($\tau_{QPR-DI}/\tau_{CPR}$) for the SF network Price graph distribution. The plots have been averaged over five randomly generated Price graph distributions with the same parameters as in Fig.~\ref{SF_price}. QPR convergence time at $\omega=0.9$ is marginally worse than CPR.}
\label{tau_price}
\end{figure}
\vspace{-0.3cm}
The SF network Barabasi-Albert graph distribution is given in Fig.~\ref{SF_barabasi}(a), and the Price graph distribution is shown in Fig.~\ref{SF_price}(a). Both have 100 vertices each and are randomly generated. Figs.~\ref{SF_barabasi}(b-c) and~\ref{SF_price}(b-c) show ranks of all the vertices pictorially for both the schemes for Barabasi Albert and Price graph distributions, respectively. We then compute the degeneracies for Barabasi-Albert and Price graph distributions in Table~\ref{SFcomp}. For the Barabasi Albert graph, CPR shows an average of 30 degeneracies, while QPR via any of the two QSW schemes shows degeneracy between 1 and 2. Similarly, for the Price graph distribution, CPR shows an average of 75 degeneracies while QPR using any of the two QSW schemes shows around ten degeneracies. For SF networks, QPR resolves degeneracies almost 30 times better than CPR in the case of Barabasi-Albert graph distribution. In contrast, in the case of Price graph distribution, it is eight times better. The plots for $\tau_{QPR}/\tau_{CPR}$ versus $\omega$ for both QSW schemes are given in Fig.~\ref{tau_BA} for Barabasi-Albert graph distribution and in Fig.~\ref{tau_price} for Price graph distribution and can be seen to be decreasing with increasing $\omega$, i.e., $\tau_{QPR}>\tau_{CPR}\ \forall\ \omega$. While checking for degeneracies for the two QSW methods, we assume $\omega=0.9$ since, at that value, the convergence time is closest to CPR (which has the minimum convergence time) while retaining quantumness. The difference between convergence time at $\omega=0.9$ and $\omega=1$, i.e., CPR, is minimal. Thus, QPR with any of the two QSW methods is remarkably better at resolving degeneracies with a marginal increase in convergence time compared to CPR.
\vspace{-0.5cm}
\subsection{Spatial networks}
\vspace{-0.5cm}
\begin{figure}[H]
\centering
\includegraphics[width=\textwidth]{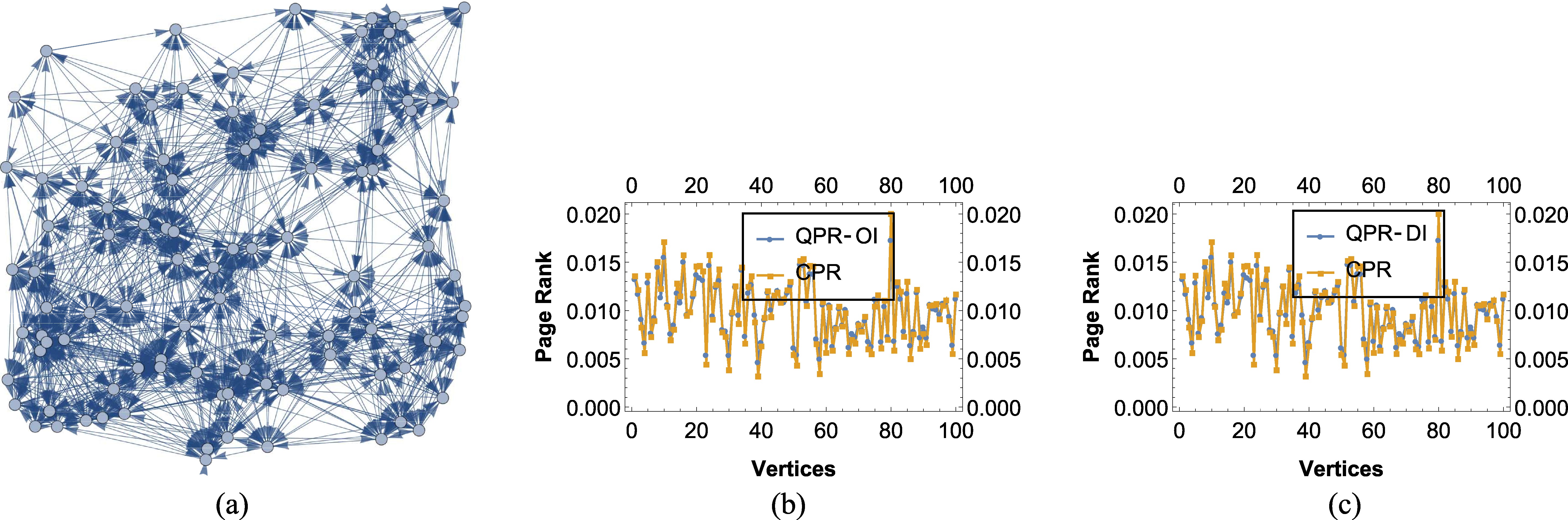}
\vspace{-0.4cm}
\caption{(a) A randomly generated spatial network with SpatialGraphDistribution[100,0.35]. SpatialGraphDistribution[n,r] in \textit{Mathematica} shows a graph of $n$ vertices uniformly distributed over a unit square. Edges exist between vertices that are at a distance at most $r$. Ranks for each vertex for QSW with (b) only incoherence (QPR-OI) and (c) dephasing with incoherence (QPR-DI) for a randomly generated spatial network with SpatialGraphDistribution[100,0.35]. CPR values are also given for comparison.}
\label{spatial_0.35}
\end{figure}
\begin{figure}[H]
\centering
\includegraphics[width=\textwidth]{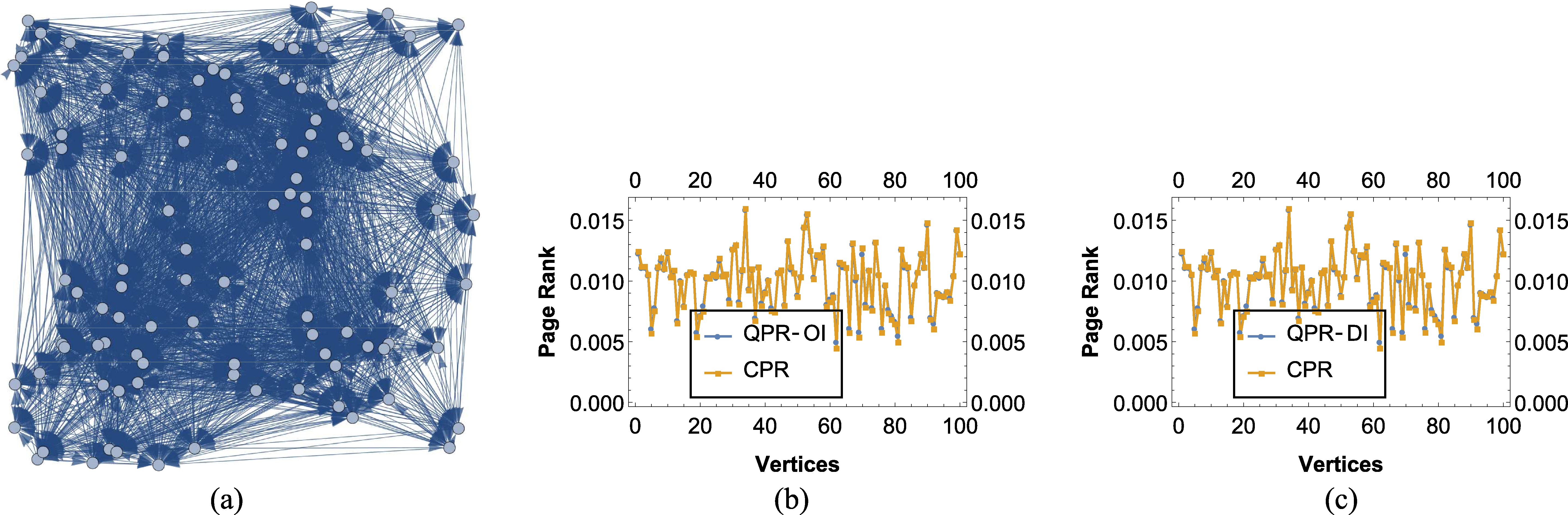}
\vspace{-0.5cm}
\caption{(a) A randomly generated spatial network with SpatialGraphDistribution[100,0.65]. Ranks for each vertex for QSW with (b) only incoherence (QPR-OI) and (c) dephasing with incoherence (QPR-DI) for a randomly generated spatial network with SpatialGraphDistribution[100,0.65]. CPR values are also given for comparison.}
\label{spatial_0.65}
\end{figure}
\vspace{-0.5cm}
\begin{table}[H]
\centering
\begin{tabular}{|cc|ccc|}
\hline\hline
\multicolumn{2}{|c|}{\multirow{4}{*}{Spatial networks}} & \multicolumn{3}{c|}{Number of degeneracies} \\ \cline{3-5}
& & \multicolumn{1}{c|}{\multirow{3}{*}{CPR}} & \multicolumn{2}{c|}{QPR} \\ \cline{4-5}
& & \multicolumn{1}{c|}{} & \multicolumn{1}{c|}{\begin{tabular}[c]{@{}c@{}}QSW (Only\\ incoherence)\end{tabular}} & \multicolumn{1}{c|}{\begin{tabular}[c]{@{}c@{}}QSW (Dephasing\\ with incoherence)\end{tabular}} \\ \hline
\multicolumn{1}{|c|}{\multirow{5}{*}{\begin{tabular}[c]{@{}c@{}}SpatialGraphDistribution[100,0.35]\end{tabular}}} & Network 1 & \multicolumn{1}{c|}{0} & \multicolumn{1}{c|}{1} & 0 \\ \cline{2-5}
\multicolumn{1}{|c|}{} & Network 2 & \multicolumn{1}{c|}{1} & \multicolumn{1}{c|}{1} & 0 \\ \cline{2-5}
\multicolumn{1}{|c|}{} & Network 3 & \multicolumn{1}{c|}{1} & \multicolumn{1}{c|}{0} & 0 \\ \cline{2-5}
\multicolumn{1}{|c|}{} & Network 4 & \multicolumn{1}{c|}{3} & \multicolumn{1}{c|}{1} & 0 \\ \cline{2-5}
\multicolumn{1}{|c|}{} & Network 5 & \multicolumn{1}{c|}{1} & \multicolumn{1}{c|}{0} & 0 \\ \hline
\multicolumn{1}{|c|}{\multirow{5}{*}{\begin{tabular}[c]{@{}c@{}}SpatialGraphDistribution[100,0.65]\end{tabular}}} & Network 1 & \multicolumn{1}{c|}{3} & \multicolumn{1}{c|}{0} & 0 \\ \cline{2-5}
\multicolumn{1}{|c|}{} & Network 2 & \multicolumn{1}{c|}{0} & \multicolumn{1}{c|}{0} & 0 \\ \cline{2-5}
\multicolumn{1}{|c|}{} & Network 3 & \multicolumn{1}{c|}{3} & \multicolumn{1}{c|}{0} & 0 \\ \cline{2-5}
\multicolumn{1}{|c|}{} & Network 4 & \multicolumn{1}{c|}{5} & \multicolumn{1}{c|}{0} & 0 \\ \cline{2-5}
\multicolumn{1}{|c|}{} & Network 5 & \multicolumn{1}{c|}{3} & \multicolumn{1}{c|}{2} & 2 \\ \hline\hline
\end{tabular}
\vspace{-0.3cm}
\caption{Comparison of the number of degeneracies in different spatial networks, with 100 vertices each- CPR versus QPR (using both QSW schemes). Five random distributions are obtained for both spatial networks.}
\label{spatialcomp}
\end{table}
\vspace{-0.7cm}
\begin{figure}[H]
\centering
\includegraphics[width=0.7\textwidth]{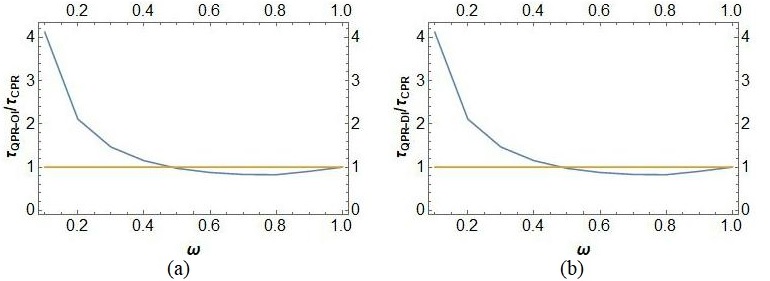}
\vspace{-0.5cm}
\caption{Convergence time $\tau_{QPR}/\tau_{CPR}$ versus $\omega$ for QSW with (a) only incoherence ($\tau_{QPR-OI}/\tau_{CPR}$) and (b) dephasing with incoherence ($\tau_{QPR-DI}/\tau_{CPR}$) for the spatial network (SpatialGraphDistribution[100,0.35]). Plots are averaged over 5 randomly generated Spatial graph distribution with $n=100,r=0.35$. $\tau_{QPR}<\tau_{CPR}$ and minimum for $\omega=0.8$.}
\label{tau_spatial0.35}
\end{figure}
\vspace{-0.7cm}
\begin{figure}[H]
\centering
\includegraphics[width=0.7\textwidth]{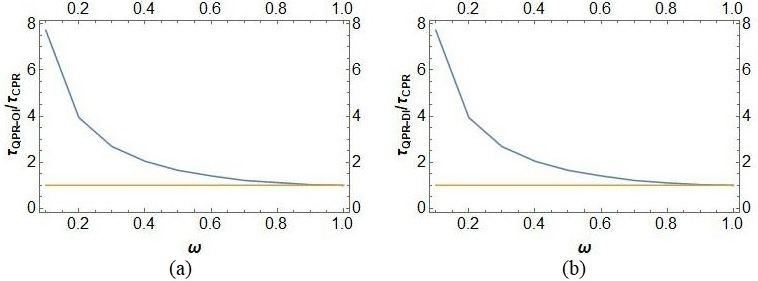}
\vspace{-0.5cm}
\caption{Convergence time $\tau_{QPR}/\tau_{CPR}$ versus $\omega$ for QSW with (a) only incoherence ($\tau_{QPR-OI}/\tau_{CPR}$) and (b) dephasing with incoherence ($\tau_{QPR-DI}/\tau_{CPR}$) for the spatial graph with SpatialGraphDistribution[100,0.65]. The plots are averaged over 5 randomly generated spatial graph distribution with $n=100,r=0.65$. $\tau_{QPR}>\tau_{CPR}\ \forall\ \omega$ with $\tau_{QPR}$ marginally worse than $\tau_{CPR}$ at $\omega=0.9$.}
\label{tau_spatial0.65}
\end{figure}
Our main aim in studying spatial networks is to see the difference in convergence time and degeneracy calculation when the distance measure is changed. Distance measure plays a vital role concerning real-world networks like communication networks or the internet. Websites are hosted at locations. While the strength of the wi-fi matters, the distance between the user's location and the website's location also matters. A website hosted at a location markedly distanced from the user will take longer to load than a website close by. As another example, take the case of walkie-talkies. The signal gets weaker and eventually cut off as the users are pulled farther apart. It is also another instance where distance measure makes a difference. The internet has a considerably higher distance measure than the communication network of walkie-talkies.
The spatial network with SpatialGraphDistribution[100,0.35] is given in Fig.~\ref{spatial_0.35}(a) and that with SpatialGraphDistribution[100,0.65] is given in Fig.~\ref{spatial_0.65}(a). Both have 100 vertices each and are randomly generated. Figs.~\ref{spatial_0.35}(b-c) and~\ref{spatial_0.65}(b-c) show ranks of all the vertices pictorially for both the schemes for SpatialGraphDistribution[100,0.35] and SpatialGraphDistribution[100,0.65] respectively. We then compute the degeneracies for both distributions in Table~\ref{spatialcomp}. For SpatialGraphDistribution[100,0.35], CPR shows an average of around a single degeneracy, while QPR via any of the two QSW schemes show almost no degeneracy. Similarly, for the SpatialGraphDistribution[100,0.65], CPR shows, on average, a degeneracy count between 2 and 3, while QPR using any of the two QSW schemes shows either no degeneracy for QPR-DI or a degeneracy between 0 and 1 for QPR-OI. The plots for $\tau_{QPR}/\tau_{CPR}$ versus $\omega$ for both QSW schemes are given in Fig.~\ref{tau_spatial0.35} for SpatialGraphDistribution[100,0.35] and in Fig.~\ref{tau_spatial0.65} for SpatialGraphDistribution[100,0.65]. The convergence time plots decrease to $\omega=0.8$ and then increase for SpatialGraphDistribution[100,0.35] while they continuously decrease with $\omega$ for SpatialGraphDistribution[100,0.65], with the convergence time at $\omega=0.9$ marginally worse than that at $\omega=1$, i.e., CPR. Thus, while checking for degeneracies using the two QSW methods, we assume $\omega=0.8$ for SpatialGraphDistribution[100,0.35] and $\omega=0.9$ for SpatialGraphDistribution[100,0.65]. We see that although the degeneracies for QPR is less than CPR for both the distance measures, the convergence time $\tau_{QPR}$ at $\omega=0.8$ with $r=0.35$ is better than $\tau_{CPR}$ while $\tau_{QPR}$ at $\omega=0.9$ with $r=0.65$ is marginally worse than $\tau_{CPR}$.

In this work, our main aim was to check for a method to improve CPR results on convergence time and resolve degeneracies. We started by explaining how page ranking works using the classical PageRank algorithm, followed by an explanation of the method employing QSW, resulting in the quantum PageRank (QPR) algorithm. We then calculated degeneracies for ER, WS, SF, and spatial complex networks using the QSW schemes. We find that QPR with only incoherence or dephasing with incoherence is significantly better than CPR for convergence times on the WS network while marginally better for the ER network, Bernoulli graph distribution, and spatial graph distribution with $r=0.35$. On the other hand, they are marginally worse for the ER network: Zachary Karate club, the two SF networks, and the spatial graph distribution with $r=0.65$.

In contrast, the improvement in resolving degeneracies by QPR is seen across the board for all networks, except for the ER network: Zachary Karate club, for which the degree of degeneracy resolution via QPR is the same as CPR. In the next section, we analyze page ranking for some smaller networks, what we expect, and check whether it is reflected in the two QSW methods and CPR. We try to understand degeneracy resolution in large networks by taking recourse to small networks, such that we can look at each vertex separately and examine its page rank. We also look into how fast we get the degeneracies resolved to find the best search algorithm among the two QSW schemes and CPR for the small networks.
% \usepackage{multirow}

% Please add the following required packages to your document %preamble:
% \usepackage{multirow}

% \usepackage{multirow}

%%%%%%%%%%%%%%%%%%%%%%%%%%%%%%%%%%%%%%%%%%%%%%%%%%%%%%% Discussion %%%%%%%%%%%%%
\section{Analysis}
In this section, we study some smaller networks to understand how the PageRank algorithm works at a smaller scale to distinguish between CPR and QPR clearly. In more extensive networks, it becomes tedious to look at the ranks of each vertex and then compare them across all methods. Thus, we rank vertices in some smaller graphs like a random eight vertex graph~\cite{sanchez2012quantum}, shown in Fig.~\ref{8vertexandgen}(a), a randomly generated spatial network of 8 vertices with $r=0.35$ and a randomly generated WS network with eight vertices and rewiring probability 0.1. We also compute the convergence time for each of these small networks.
\vspace{-0.2cm}
\subsection{Random 8 vertex graph}
\vspace{-0.5cm}
%We rank the random vertex graph from Ref.~\cite{sanchez2012quantum} using CPR and the two QSW methods to see which is better.
\begin{figure}[H]
\centering
\includegraphics[width=0.7\textwidth]{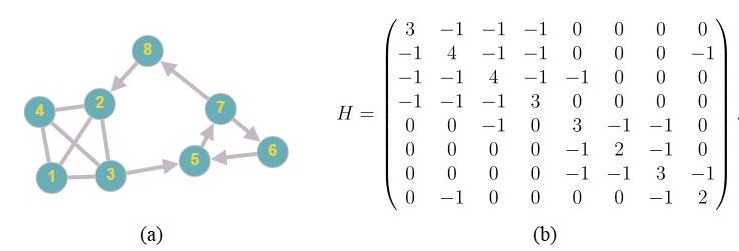}
\vspace{-0.4cm}
\caption{(a) A random 8 vertex graph~\cite{sanchez2012quantum}. and (b) the generator matrix for the same.}
\label{8vertexandgen}
\end{figure}
\vspace{-0.5cm}
\begin{figure}[H]
\centering
\includegraphics[width=0.7\textwidth]{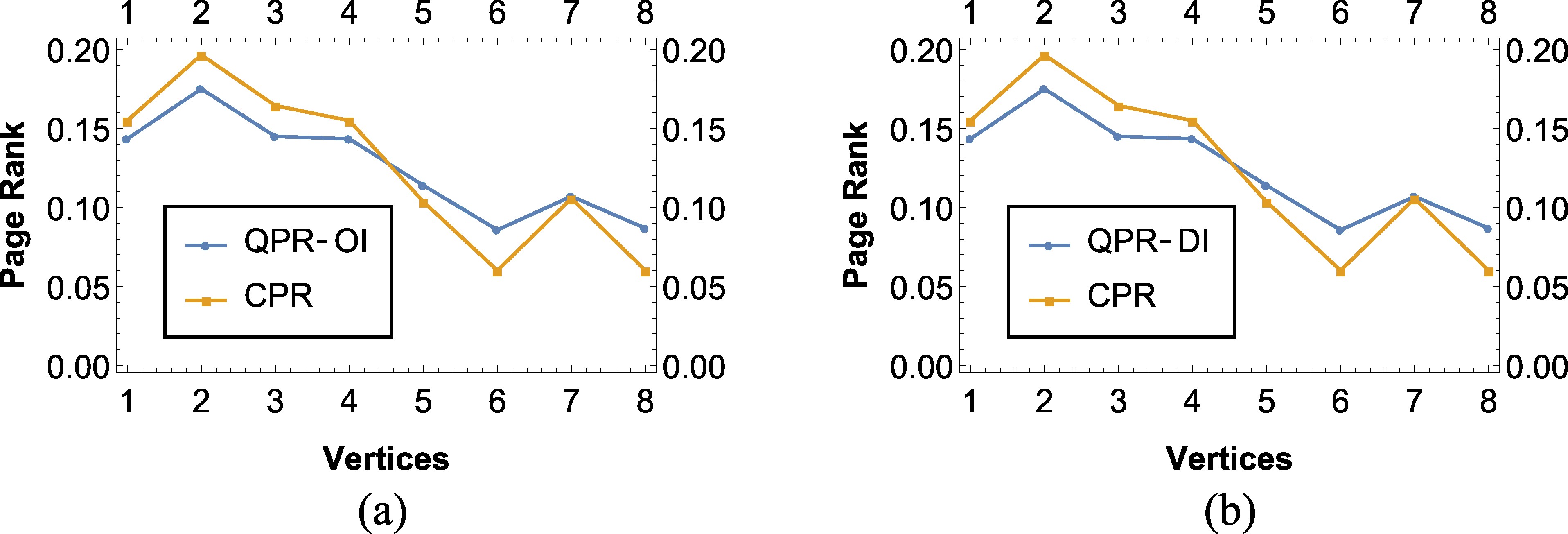}
\vspace{-0.4cm}
\caption{Ranks for each vertex for a random eight vertex graph for QSW with (a) only incoherence (QPR-OI) and (b) dephasing with incoherence (QPR-DI). CPR values are also given for comparison.}
\label{QPR_sanchez}
\end{figure}
\begin{figure}[H]
\centering
\includegraphics[width=0.7\textwidth]{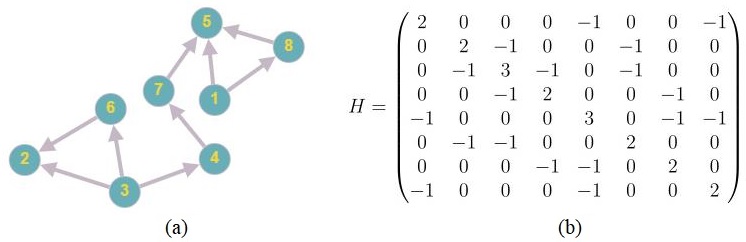}
\vspace{-0.4cm}
\caption{(a) A randomly generated spatial network of 8 vertices and $r=0.35$ and (b) the generator matrix for the same.}
\label{Spatial8vertexandgen}
%\end{adjustbox}
\end{figure}

\vspace{-0.5cm}
\begin{table}[H]
\centering
\begin{tabularx}{\textwidth}{|Y|Y|Y|Y|}
\hline\hline
& \multicolumn{3}{c|}{Page rank} \\ \cline{2-4}
\multirow{-2}{*}{\begin{tabular}[c]{@{}c@{}}Random 8\\vertex graph\end{tabular}} & \multicolumn{1}{c|}{} & \multicolumn{2}{c|}{QPR} \\ \cline{1-1} \cline{3-4}
Vertex & \multicolumn{1}{c|}{\multirow{-2}{*}{CPR}} & \multicolumn{1}{c|}{\begin{tabular}[c]{@{}c@{}}QSW (Only\\ incoherence)\end{tabular}} & \begin{tabular}[c]{@{}c@{}}QSW (Dephasing \\with incoherence)\end{tabular} \\ \hline
2 & 0.1965 & 0.1750 & 0.1750\\
3 & 0.1644 & 0.1449 & 0.1449\\
1 & \cellcolor{ashgrey}0.1549 & \cellcolor{ashgrey}0.1434 & \cellcolor{ashgrey}0.1434\\
4 & \cellcolor{ashgrey}0.1549 & \cellcolor{ashgrey}0.1434 & \cellcolor{ashgrey}0.1434\\
5 & 0.1035 & 0.1139 & 0.1139\\
7 & 0.1057 & 0.1069 & 0.1070\\
8 & \cellcolor{khaki(x11)(lightkhaki)}0.0601 & 0.0868 & 0.0868\\
6 & \cellcolor{khaki(x11)(lightkhaki)}0.0601 & 0.0857 & 0.0857\\
\hline\hline
\end{tabularx}
\vspace{-0.3cm}
\caption{QPR, CPR comparison for a random 8 vertex graph (Fig.~\ref{8vertexandgen}(a)), QSW with only incoherence (Eq.~(\ref{incoh})) and dephasing with incoherence (Eq.~(\ref{dep-incoh})).}
\label{QPRcompsanchez}
\end{table}
\vspace{-0.5cm}
\begin{figure}[H]
\centering
\includegraphics[width=0.7\textwidth]{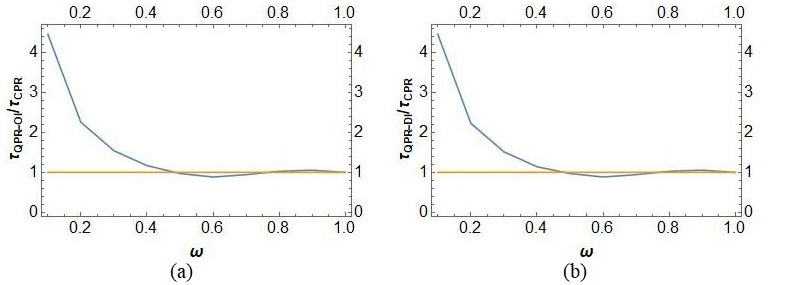}
\vspace{-0.4cm}
\caption{Convergence time $\tau_{QPR}/\tau_{CPR}$ versus $\omega$ for QSW with (a) only incoherence ($\tau_{QPR-OI}/\tau_{CPR}$) and (b) dephasing with incoherence ($\tau_{QPR-DI}/\tau_{CPR}$) for a random 8 vertex graph (see Fig.~\ref{8vertexandgen}(a)).}
\label{tau_sanchez}
\end{figure}
\vspace{-0.5cm}
We compare the ranks of all the vertices using CPR and QPR for both QSW schemes for a random eight vertex graph. Fig.~\ref{8vertexandgen} depicts the random eight vertex graph and the corresponding generator matrix. Fig.~\ref{QPR_sanchez} depicts the page rank of each vertex using QSW for both the schemes. From Table~\ref{QPRcompsanchez} we find that CPR shows degeneracies between vertices $1-4$ and $6-8$, while QPR with only incoherence and dephasing with incoherence schemes resolve degeneracies between vertices 6-8 while that for 1-4 remains. Thus, even though QPR is better than CPR, it is not $100\%$ efficient; some degeneracies are still there. From Fig.~\ref{8vertexandgen}, we expect that vertex two should have the highest rank due to the maximum number of links, while vertices 8 and 6 should have the lowest ranks due to them having the least number of links. The other vertices should have intermediate rankings. Both CPR and QSW methods of only incoherence and dephasing with incoherence reflect this expectation. The plots for $\tau_{QPR}/\tau_{CPR}$ versus $\omega$ are given in Fig.~\ref{tau_sanchez}. The curves decrease and then increase again, achieving a minimum of $\omega=0.6$. Thus, for ranking via the two QSW schemes, we use $\omega=0.6$ since the convergence time is minimum at that value of $\omega$. Thus, we see that QPR via any of the two QSW methods resolves degeneracies better and faster than CPR for an eight vertex graph. Now, we compare our results with those from Ref.~\cite{sanchez2012quantum}, which uses the QSW approach of dephasing with incoherence for ranking the eight vertex graphs. Our method's ranks are slightly different from those given in Ref.~\cite{sanchez2012quantum}. The reason behind this is because we use the generator matrix provided in Fig.~\ref{8vertexandgen}(b), while Ref.~\cite{sanchez2012quantum} uses the generator matrix with entry $H_{ij}=1$ if $i,j$ are connected and zero otherwise.
\vspace{-0.5cm}
\subsection{Random spatial network with 8 vertices (SpatialGraphDistribution[8,0.35]), see also Figs.~\ref{spatial_0.35} and~\ref{Spatial8vertexandgen}}
\vspace{-0.3cm}
We generate a random spatial network of 8 vertices with $r=0.35$, discuss what ranking we expect for this small network, and then verify whether our expectations are reflected in the two QSW methods and CPR and in what convergence time.
\vspace{-0.5cm}
\begin{figure}[H]
\centering
\includegraphics[width=0.7\textwidth]{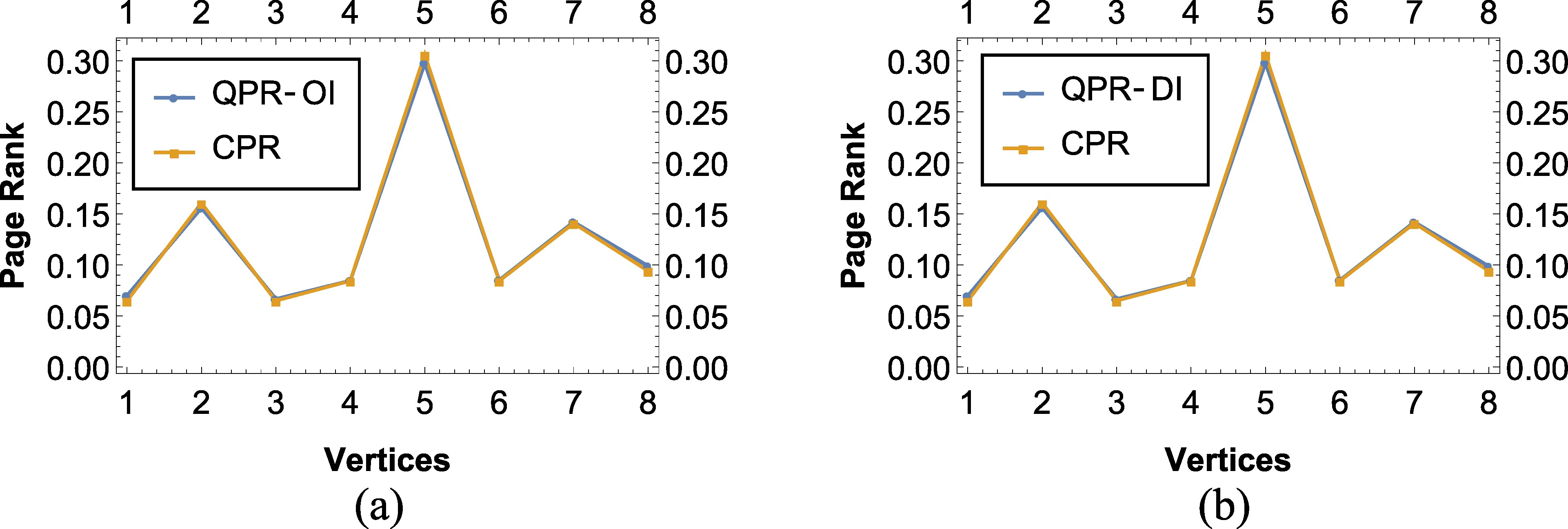}
\vspace{-0.4cm}
\caption{Ranks for each vertex for QSW with (a) only incoherence (QPR-OI) and (b) dephasing with incoherence (QPR-DI) for the randomly generated eight vertex spatial network in Fig.~\ref{Spatial8vertexandgen}(a). CPR values are also given for comparison.}
\label{QPR_spatial8vertex}
\end{figure}
\vspace{-0.2cm}
\begin{table}[H]
\centering
\begin{tabularx}{\textwidth}{|Y|Y|Y|Y|}
\hline\hline
& \multicolumn{3}{c|}{Page rank} \\ \cline{2-4}
\multirow{-2}{*}{\begin{tabular}[c]{@{}c@{}}Random spatial network\\with 8 vertices\end{tabular}} & \multicolumn{1}{c|}{} & \multicolumn{2}{c|}{QPR} \\ \cline{1-1} \cline{3-4}
Vertex & \multicolumn{1}{c|}{\multirow{-2}{*}{CPR}} & \multicolumn{1}{c|}{\begin{tabular}[c]{@{}c@{}}QSW (Only\\ incoherence)\end{tabular}} & \begin{tabular}[c]{@{}c@{}}QSW (Dephasing \\with incoherence)\end{tabular} \\ \hline
5 & 0.3058 & 0.2976 & 0.2980\\
2 & 0.1604 & 0.1559 & 0.1561\\
7 & 0.1409 & 0.1419 & 0.1418\\
8 & 0.0942 & 0.0989 & 0.0986\\
6 & \cellcolor{ashgrey}0.0844 & 0.0850 & 0.0849\\
4 & \cellcolor{ashgrey}0.0844 & 0.0845 & 0.0845\\
1 & \cellcolor{khaki(x11)(lightkhaki)}0.0649 & 0.0697 & 0.0695\\
3 & \cellcolor{khaki(x11)(lightkhaki)}0.0649 & 0.0665 & 0.0665\\
\hline\hline
\end{tabularx}
\vspace{-0.3cm}
\caption{Page rank CPR versus QPR (for both the QSW schemes) for the random spatial network with eight vertices (see Fig.~\ref{Spatial8vertexandgen}(a)). Degeneracies in CPR are resolved in both QPR schemes.}
\label{spatial8vertexrankcomp}
\end{table}
\vspace{-0.2cm}
\begin{figure}[H]
\centering
\includegraphics[width=0.7\textwidth]{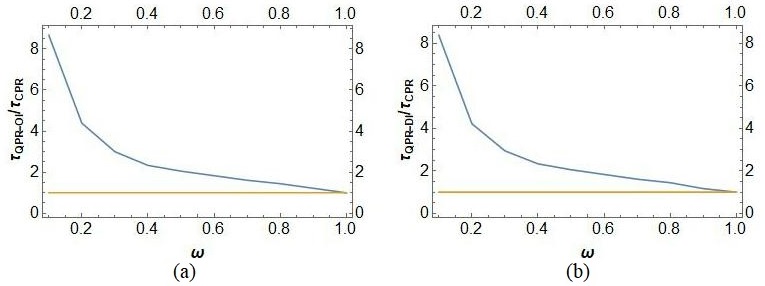}
\caption{Convergence time $\tau_{QPR}/\tau_{CPR}$ versus $\omega$ for QSW with (a) only incoherence ($\tau_{QPR-OI}/\tau_{CPR}$) and (b) dephasing with incoherence ($\tau_{QPR-DI}/\tau_{CPR}$) for the randomly generated spatial network of 8 vertices as in Fig.~\ref{Spatial8vertexandgen}(a). Convergence time in QPR schemes is always larger than CPR irrespective of $\omega$.}
\label{tau_spatial_8vertex}
\end{figure}
In Fig.~\ref{Spatial8vertexandgen}(a), we can see the randomly generated spatial network with eight vertices and $r=0.35$ and its corresponding generator matrix.
Fig.~\ref{QPR_spatial8vertex} depicts the ranks of each vertex for both of the QSW schemes. We see degeneracies between vertices 1-3 and 4-6 using CPR for the randomly generated spatial network. On the other hand, employing QPR via any of the two QSW schemes resolves degeneracies between all the vertices (see Table~\ref{spatial8vertexrankcomp}). From Fig.~\ref{Spatial8vertexandgen}, we can see that vertex 5 has three incoming links and none outgoing. Thus, it makes sense that vertex 5 has the highest rank using both CPR and the two QSW methods. Similarly, vertex 2 has two incoming links and none outgoing, so it receives the second highest rank using CPR. QPR Vertices 1 and 3 have two and three outgoing links, respectively, and none incoming. Consequently, they will have the lowest ranks. Since vertex 3 has a higher number of outgoing links, it is plausible to assume that it will have a lower rank than vertex 1. CPR, however, allots both of them the same rank while QPR allots vertex 1, a higher rank than 3, which is in line with our expectation. Now, vertices 4, 6, 7, and 8 have one outgoing and one incoming link each. Be that as it may, vertices 4 and 6 have links from vertex 3, which has the lowest rank. Similarly, vertex eight is linked to vertex 1, which has the second lowest rank. Thus, vertices 4 and 6 get the lowest ranks among the four while vertex 7 gets the highest, and vertex eight ranks in the middle by CPR as well as QPR CPR is not able to distinguish between vertices 4 and 6, giving them the same ranks, but QPR can set the two apart and gives vertex six a higher rank than 4. The logic here is that vertex 6 has a link to vertex two while vertex 4 has a link to 7. Vertex 2 is ranked higher than 7; thus, vertex six is ranked higher than four by QPR. Here, it becomes apparent that QPR can resolve degeneracies better than CPR, and the ranks allotted also do not defy expectations. The plots for $\tau_{QPR}/\tau_{CPR}$ versus $\omega$ are given in Fig.~\ref{tau_spatial_8vertex}. We find $\tau_{QPR}$ is always greater than $\tau_{CPR}$ irrespective of $\omega$ and the QSW scheme. It suggests that although QPR best resolves degeneracies, it takes some more time, around 1.25 times for $\omega=0.9$. Thus, we have chosen $\omega=0.9$ for ranking via the two QSW schemes. The reason is that for this value of $\omega$, we have a convergence time closest to the minimum (which is at $\omega=1$, i.e., CPR) and a better degeneracy resolution than CPR.
\vspace{-0.3cm}
\subsection{Random WS network with eight vertices}
\vspace{-0.1cm}
Like the previous two cases, we analyze this network, predict the ranking we expect, and check to what extent our expectations are realized. We also calculate the convergence for this network.
\vspace{-0.2cm}
\begin{figure}[H]
\centering
\includegraphics[width=0.7\textwidth]{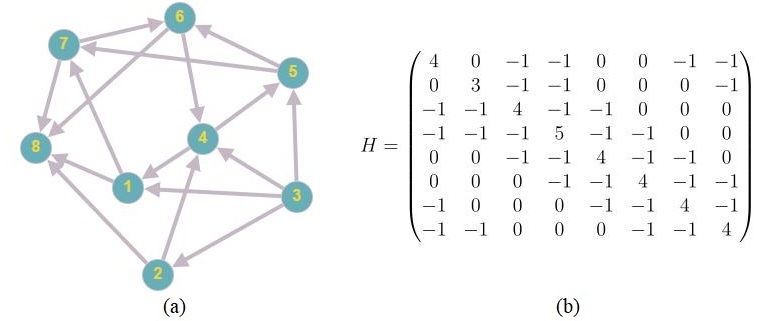}
\vspace{-0.4cm}
\caption{(a) A randomly generated WS network of 8 vertices with rewiring probability 0.1, i.e., WattsStrogatzGraphDistribution[8, 0.1]. (b) The generator matrix for the same.}
\label{WS8vertexandgen}
\end{figure}
\vspace{-0.5cm}
\begin{figure}[H]
\centering
\includegraphics[width=0.7\textwidth]{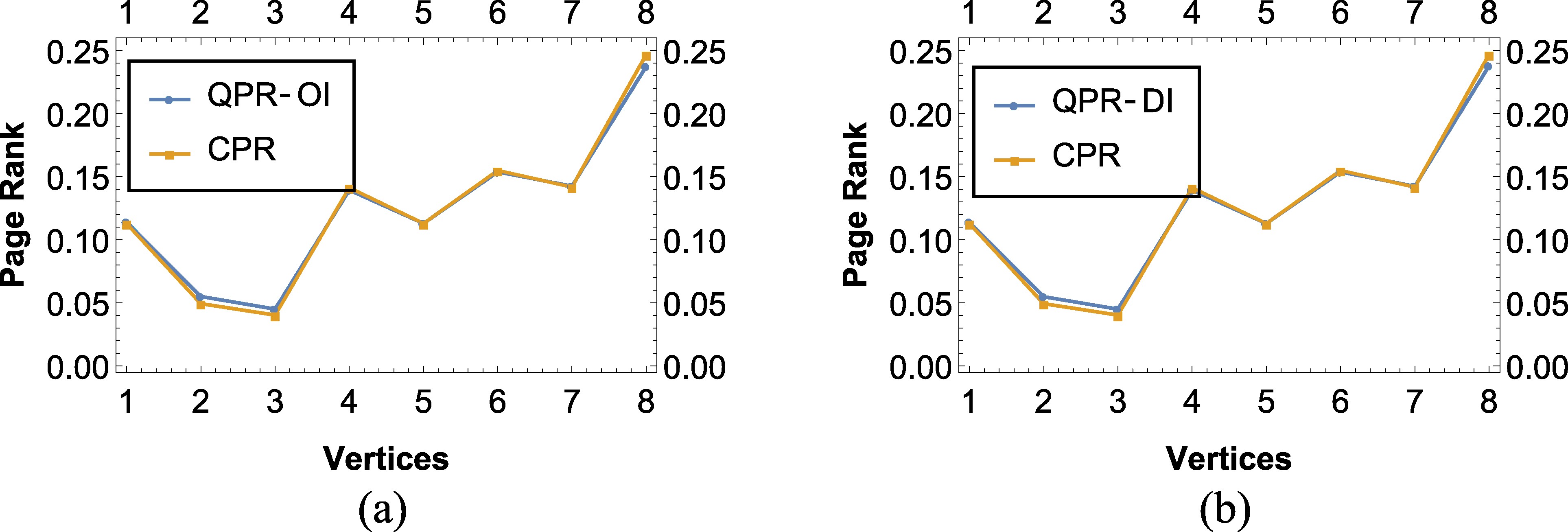}
\vspace{-0.4cm}
\caption{Ranks for each vertex for QSW with (a) only incoherence (QPR-OI) and (b) dephasing with incoherence (QPR-DI) for the randomly generated WS network with parameter same as in Fig.~\ref{WS8vertexandgen}(a). CPR values are also given for comparison.}
\label{QPR_WS8}
\end{figure}
\vspace{-0.5cm}
\begin{table}[H]
\centering
\begin{tabularx}{\textwidth}{|Y|Y|Y|Y|}
\hline\hline
& \multicolumn{3}{c|}{Page rank} \\ \cline{2-4}
\multirow{-2}{*}{\begin{tabular}[c]{@{}c@{}}Random WS network\\with 8 vertices\end{tabular}} & \multicolumn{1}{c|}{} & \multicolumn{2}{c|}{QPR} \\ \cline{1-1} \cline{3-4}
Vertex & \multicolumn{1}{c|}{\multirow{-2}{*}{CPR}} & \multicolumn{1}{c|}{\begin{tabular}[c]{@{}c@{}}QSW (Only\\ incoherence)\end{tabular}} & \begin{tabular}[c]{@{}c@{}}QSW (Dephasing \\with incoherence)\end{tabular} \\ \hline
8 & 0.2468 & 0.2374 & 0.2378\\
6 & 0.1549 & 0.1538 & 0.1538\\
7 & 0.1418 & 0.1425 & 0.1424\\
4 & 0.1412 & 0.1394 & 0.1393\\
1 & \cellcolor{ashgrey}0.1129 & 0.1140 & 0.1139\\
5 & \cellcolor{ashgrey}0.1129 & 0.1129 & 0.1129\\
2 & 0.0493 & 0.0550 & 0.0548\\
3 & 0.0403 & 0.0450 & 0.0450\\
\hline\hline
\end{tabularx}
\vspace{-0.3cm}
\caption{Page rank CPR versus QPR (using the two QSW schemes) for the randomly generated WS network with parameters same as in Fig.~\ref{WS8vertexandgen}(a). QPR via both schemes resolves the degeneracy in CPR.}
\label{WS8rankcomp}
\end{table}
\begin{figure}[H]
\centering
\includegraphics[width=0.7\textwidth]{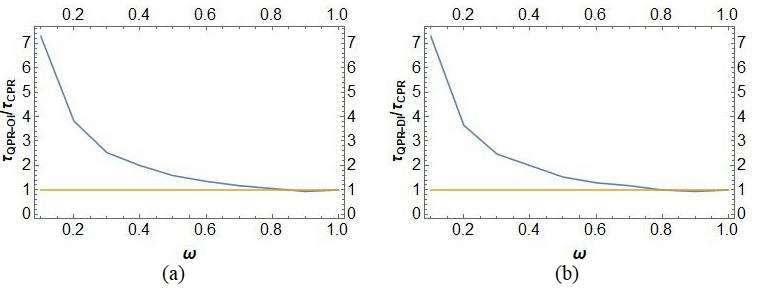}
\caption{Convergence time $\tau_{QPR}/\tau_{CPR}$ versus $\omega$ for QSW with (a) only incoherence ($\tau_{QPR-OI}/\tau_{CPR}$) and (b) dephasing with incoherence ($\tau_{QPR-DI}/\tau_{CPR}$) for the randomly generated WS network of 8 vertices in Fig.~\ref{WS8vertexandgen}(a). $\tau_{QPR}$ at $\omega=0.9$ is slightly less than $\tau_{CPR}$.}
\label{tau_WS8}
\end{figure}
Fig.~\ref{WS8vertexandgen} depicts the randomly generated WS network of 8 vertices and rewiring probability $p=0.1$ and its generator matrix.
Fig.~\ref{QPR_WS8} depicts the ranks of each of the vertices for QPR using both the QSW schemes. The comparison between ranks assigned by CPR and QPR using the two QSW methods is made in Table~\ref{WS8rankcomp}. When ranking via CPR, degeneracies exist between vertices 1 and 5, which QPR using both the QSW schemes, manages to resolve. From Fig.~\ref{WS8vertexandgen}, vertex 8 has four incoming links and none outgoing, while vertex 3 has four outgoing links and none incoming. Hence, they are ranked the highest and lowest respectively by both CPR and QPR Vertex 2 has one incoming link from vertex three and two outgoing links. Thus, vertex 2 has the second lowest rank among all vertices. Vertices 4, 6, and 7 have intermediate ranks, with vertex 4 ranking the lowest among the three because of having a link from vertex 3. Vertex 6 ranks higher than vertex seven and highest among vertices 4, 6, and 7 as it has an incoming link from vertex 7. CPR allots vertices 1 and 5 the same rank. QPR separates the two as vertex 1 has a link to vertex eight, which has the highest rank, while vertex five does not have this link. Thus, QPR allots vertex one a higher rank than vertex 5. For this case, we can also conclude that QPR resolves degeneracies better than CPR, and the ranks allotted are within our expectations. The plots for $\tau_{QPR}/\tau_{CPR}$ versus $\omega$ are given in Fig.~\ref{tau_WS8} and at $\omega=0.9$, $\tau_{QPR}<\tau_{CPR}$. We use $\omega=0.9$ while checking for degeneracies via the two QSW schemes as convergence time for this $\omega$ is minimum and resolves degeneracies better than CPR. Thus, for this case, QPR not only resolves all the degeneracies but also has a better convergence time than CPR.\\

In this section, we analyzed the vertex ranks for the small networks like the random eight vertex graph, the randomly generated spatial network, and the WS network with eight vertices so that the difference between CPR and QPR becomes apparent. We also checked whether CPR and QPR methods follow our expectations when ranking vertices. Through this, we can verify that the two QPR methods return the correct ranks, which was impractical to check while dealing with bigger networks. We found that QPR resolves degeneracies better than CPR but is not perfect, i.e., it cannot resolve all degeneracies, and some still exist. QPR trumps CPR in degeneracy resolution and convergence time for the random eight vertex graph and a random WS network with eight vertices. For the random spatial network with eight vertices, although QPR resolves all the
degeneracies, $\tau_{QPR}$ is marginally worse than $\tau_{CPR}$ for all $\omega$. Overall, QPR is still better than CPR for these small networks.
\section{Conclusion and Future plans}
In conclusion, page ranking using QSW via any of the two schemes is a valuable tool compared to the original CPR algorithm. We introduced the two QSW schemes of only incoherence and dephasing with incoherence, after which we applied said schemes in the PageRank algorithm to obtain the quantum PageRank algorithm (QPR). We aimed to check whether QPR can resolve degeneracies unresolvable by CPR and the number of navigation steps required to reach the stationary state, which outputs the ranks of vertices in a graph. We utilized a method best suited to solving page ranking degeneracies, i.e., any of the two: QSW with only incoherence or dephasing with incoherence scheme, as their efficacy of resolving the degeneracies, is, in general, higher than CPR and the convergence time either marginally differing or significantly better than CPR for all networks. In ER network: Zachary Karate club, the convergence time is marginally worse than CPR, while the degree of degeneracy resolution is identical for both QPR and CPR. For the ER network: Bernoulli graph distribution, no degeneracies exist in CPR and QPR, while convergence time for QPR is marginally better than CPR.

In contrast, for the WS network, the improvement in both degeneracy resolution and convergence time is often better for QPR than CPR. For SF networks: Barabasi-Albert graph distribution and Price graph distribution, degeneracy resolution via QPR is much better while convergence time is marginally worse than CPR. For spatial networks with distance measure $r=0.35$, we see that degeneracy resolution at $\omega=0.8$ via QPR is significantly better than CPR, while the convergence time at $\omega=0.8$ is slightly better than CPR. On the other hand, for the spatial network with distance measure $r=0.65$, the degeneracy resolution at $\omega=0.9$ is better than CPR, but convergence time at $\omega=0.9$ is marginally worse than CPR. Thus, we find that changing the distance measure makes the convergence time change distinctly. The best result for QPR is seen for the WS network, where the degeneracy resolution and convergence time are significantly better than CPR. In SF networks, too, although the convergence time is marginally worse for QPR, the degeneracy resolution is multiple times better than CPR Ergo; we can safely say that QPR via any of the two QSW schemes is a better choice than CPR for the PageRank algorithm.

QPR can be readily implemented without the need of any quantum hardware, as of today what are available in the quantum realm are NISQ quantum computers which anyway function with errors. Our paper exploits new features that quantum rules bring out when ranking complex networks on a classical computer. As we show in the Mathematica code in Appendix, we show the difference when running the our code with quantum rules on a classical computer wherein our code does beat a classical page rank algorithm working in the same classical computer in the case of the Watts-Strogatz network as far as convergence time is concerned. Furthermore, when we consider degeneracy, then QPR is always better than CPR even when working on a classical computer. 

However, the possible implementation of QPR, in a quantum machine will open the path to a more efficient application of quantum rank techniques. To this aim and considering the recent advances in quantum annealing protocols, one should address the optimal mapping of the network nodes into the (minimal) number of qubits, considering the complex network topology of the interactions between them. One can do a one to one between vertices of our networks and qubits of a quantum computer which would mean that our algorithm if one wants to run a quantum machine will need $100$ qubits  or given that qubits can be entangled, a question to ponder can be, Will it be possible to map to a lesser number of qubits but which show the same complexity as the $100$ vertex graph in question? Perhaps answers to such questions may be the focus of future works  on implementing QPR in quantum computers.

Quantum internet~\cite{kimble2008quantum} is still in the process of development but once done, it will require quantum algorithms to administer. In this work, we have provided a quantum algorithm and implemented it to solve a classical problem involving complex networks. We have implemented QSW in ER, WS, SF, and spatial networks. In the future, more types of networks, e.g., neural networks, can be included. Neural networks (NN) are interconnected circuits of neurons, implementing algorithms to analyze a set of data, much like how a human brain works. They are used for computing. There are two types of NN, a biological neural network (BNN), e.g., the human brain, and an artificial neural network (ANN), making use of artificial intelligence (AI). Neural networks use parallel computing which leads to faster computation. Researchers have been trying to obtain an even faster computation by combining the unitary dynamics of quantum mechanics with the dynamics of neural networks, obtaining a quantum neural network (QNN). However, due to the irreversible dynamics of neural networks, combining CTQW with the same has proven to be challenging since CTQW requires unitarity and demands reversibility.
Future directions include implementing quantum stochastic walks on neural networks since QSW can incorporate both reversible and irreversible dynamics~\cite{schuld2014quantum}.
\section{Acknowledgements}
CB would like to thank the Science and Engineering Research Board (SERB) for funding under MATRICS grant ``Nash equilibrium versus Pareto optimality in N-Player games" (MTR/2018/000070) and also for funding under CRG grant ``Josephson junctions with strained Dirac materials and their application in quantum information processing" (CRG/2019/006258).
%%%%%%%%%%%%%%%%%%%%%%%%%%%%%%%%%%%%%%%%%%%%%%%%%%%%%%%%%%%%%%%%%%%%%%%%%%%55
\section{Appendix}

\subsection{Code for plotting page rank versus vertices}
Below we provide the \textit{Mathematica} code for plotting page rank versus vertices. It has been used to generate Fig.~\ref{ER_watts} for the WS network. For the other networks, changes can be made accordingly.

\begin{widetext}
\begin{lstlisting}
<< QSWalk`
Remove["Global`*"]
G = DirectedGraph[
RandomGraph[WattsStrogatzGraphDistribution[100, 0.2]], "Random"]
alpha = 0.9;
n = VertexCount[G];
H = GeneratorMatrix[UndirectedGraph[G], 1];
list1 = LindbladSet[GoogleMatrix[G, alpha]];
list2 = Table[
SparseArray[{{i, i} ->
Sqrt[Abs[GoogleMatrix[G, alpha][[i, i]]]]}, {n, n}], {i, n}];
LkSetdec = DeleteCases[list1, Alternatives @@ list2];
LkSetdephanddec = LindbladSet[ GoogleMatrix[G, alpha]];
omega = 0.4; tf = 200.;
rho0 = SparseArray[{{i_, i_} -> 1/n}, {n, n}, 0];
QPRdec = QuantumStochasticWalk[H, LkSetdec, omega, rho0, tf] //
Diagonal // Chop;
QPRdephanddec =
QuantumStochasticWalk[H, LkSetdephanddec, omega, rho0, tf] //
Diagonal // Chop;
CPR = PageRankCentrality[G, alpha];
QPRdecac = SetPrecision[QPRdec, 4];
QPRdephanddecac = SetPrecision[QPRdephanddec, 4];
CPRac = SetPrecision[CPR, 4];
QPRdecdeg = CountDistinct[QPRdecac]
QPRdephanddecdeg = CountDistinct[QPRdephanddecac]
CPRdeg = CountDistinct[CPRac]

ListLinePlot[{QPRdec, CPR}, PlotRange -> Full,
PlotMarkers -> Automatic,
Axes -> False, Frame -> True,
FrameLabel -> {{Style["Page Rank", Bold, Black, 16],
None}, {Style["Vertices", Black, Bold, 16], None}},
LabelStyle -> Directive[Black, FontSize -> 16], FrameTicks -> All,
PlotLegends ->
Placed[
LineLegend[{"QPR-OI", "CPR"}, LegendFunction -> Framed], {0.5,
0.8}]]

ListLinePlot[{QPRdephanddec, CPR}, PlotRange -> Full,
PlotMarkers -> Automatic,
Axes -> False, Frame -> True,
FrameLabel -> {{Style["Page Rank", Bold, Black, 16],
None}, {Style["Vertices", Black, Bold, 16], None}},
LabelStyle -> Directive[Black, FontSize -> 16], FrameTicks -> All,
PlotLegends ->
Placed[
LineLegend[{"QPR-DI", "CPR"}, LegendFunction -> Framed], {0.5,
0.8}]]
\end{lstlisting}

\end{widetext}

\subsection{Code for plotting convergence time}
We have presented the \textit{Mathematica} code for plotting convergence time versus $\omega$. This code is to generate Fig.~\ref{tau_WS}, i.e., for the WS network. This code makes use of \textit{QSWalk} package~\cite{QSWalk,QSWalkdownload}. For other networks, changes can be made accordingly.

\begin{widetext}
\begin{lstlisting}
<< QSWalk`
Remove["Global`*"]
G = DirectedGraph[
RandomGraph[WattsStrogatzGraphDistribution[100, 0.2]], "Random"]
alpha = 0.9;
n = VertexCount[G];
H = GeneratorMatrix[UndirectedGraph[G]];
list1 = LindbladSet[GoogleMatrix[G, alpha]];
list2 = Table[
SparseArray[{{i, i} ->
Sqrt[Abs[GoogleMatrix[G, alpha][[i, i]]]]}, {n, n}], {i, n}];
LkSetdec = DeleteCases[list1, Alternatives @@ list2];
LkSetdephanddec = LindbladSet[GoogleMatrix[G, alpha]]; tf = 800.;
rho0 = SparseArray[{{i_, i_} -> 1/n}, {n, n}, 0];
tsteplistdec = List[];
tsteplistdephanddec = List[];
omega = 0.1;
While[omega <= 1.,
QPRdec =
QuantumStochasticWalk[H, LkSetdec, omega, rho0, tf] // Diagonal //
Chop;
QPRdephanddec =
QuantumStochasticWalk[H, LkSetdephanddec, omega, rho0, tf] //
Diagonal // Chop;
CPR = PageRankCentrality[G, alpha];
rhodec = rho0 // Diagonal // Chop;
normdec[rhodec_] := Sqrt[Sum[(rhodec[[i]] - QPRdec[[i]])^2, {i, n}]];
normdephanddec[rhodephanddec_] :=
Sqrt[Sum[(rhodephanddec[[i]] - QPRdephanddec[[i]])^2, {i, n}]];
tol = 10^(-6);
QSWdec = List[];
tstepdec = 0;
While[normdec[rhodec] >= tol,
rhodec =
QuantumStochasticWalk[H, LkSetdec, omega, rho0, tstepdec] //
Diagonal // Chop; tstepdec++];
tsteplistdec = Append[tsteplistdec, tstepdec];
rhodephanddec = rho0 // Diagonal // Chop;
QSWdephanddec = List[];
tstepdephanddec = 0;
While[normdephanddec[rhodephanddec] >= tol,
rhodephanddec =
QuantumStochasticWalk[H, LkSetdephanddec, omega, rho0,
tstepdephanddec] // Diagonal // Chop; tstepdephanddec++];
tsteplistdephanddec = Append[tsteplistdephanddec, tstepdephanddec];
omega = omega + 0.1];
taudecratio = tsteplistdec/tsteplistdephanddec[[-1]];
taudephanddecratio = tsteplistdephanddec/tsteplistdephanddec[[-1]];
omegaval = Range[0.1, 1, 0.1];
ListPlot[Transpose@{omegaval, taudecratio}, PlotRange -> All,
Joined -> True,
Axes -> False, Frame -> True ,
FrameLabel -> {{Style[
"\!\(\*SubscriptBox[\(\[Tau]\), \(QPR - \
OI\)]\)/\!\(\*SubscriptBox[\(\[Tau]\), \(CPR\)]\)", Black, 16, Bold],
None}, {Style["\[Omega]", Black, 16, Bold], None}},
LabelStyle -> Directive[Black, FontSize -> 16], FrameTicks -> All]
ListPlot[Transpose@{omegaval, taudephanddecratio}, PlotRange -> All,
Joined -> True,
Axes -> False, Frame -> True ,
FrameLabel -> {{Style[
"\!\(\*SubscriptBox[\(\[Tau]\), \(QPR - \
DI\)]\)/\!\(\*SubscriptBox[\(\[Tau]\), \(CPR\)]\)", Black, 16, Bold],
None}, {Style["\[Omega]", Black, 16, Bold], None}},
LabelStyle -> Directive[Black, FontSize -> 16], FrameTicks -> All]

\end{lstlisting}
\end{widetext}
\bibliography{ref.bib}

\begin{thebibliography}{36}%
\makeatletter
\providecommand \@ifxundefined [1]{%
 \@ifx{#1\undefined}
}%
\providecommand \@ifnum [1]{%
 \ifnum #1\expandafter \@firstoftwo
 \else \expandafter \@secondoftwo
 \fi
}%
\providecommand \@ifx [1]{%
 \ifx #1\expandafter \@firstoftwo
 \else \expandafter \@secondoftwo
 \fi
}%
\providecommand \natexlab [1]{#1}%
\providecommand \enquote  [1]{``#1''}%
\providecommand \bibnamefont  [1]{#1}%
\providecommand \bibfnamefont [1]{#1}%
\providecommand \citenamefont [1]{#1}%
\providecommand \href@noop [0]{\@secondoftwo}%
\providecommand \href [0]{\begingroup \@sanitize@url \@href}%
\providecommand \@href[1]{\@@startlink{#1}\@@href}%
\providecommand \@@href[1]{\endgroup#1\@@endlink}%
\providecommand \@sanitize@url [0]{\catcode `\\12\catcode `\$12\catcode
  `\&12\catcode `\#12\catcode `\^12\catcode `\_12\catcode `\%12\relax}%
\providecommand \@@startlink[1]{}%
\providecommand \@@endlink[0]{}%
\providecommand \url  [0]{\begingroup\@sanitize@url \@url }%
\providecommand \@url [1]{\endgroup\@href {#1}{\urlprefix }}%
\providecommand \urlprefix  [0]{URL }%
\providecommand \Eprint [0]{\href }%
\providecommand \doibase [0]{https://doi.org/}%
\providecommand \selectlanguage [0]{\@gobble}%
\providecommand \bibinfo  [0]{\@secondoftwo}%
\providecommand \bibfield  [0]{\@secondoftwo}%
\providecommand \translation [1]{[#1]}%
\providecommand \BibitemOpen [0]{}%
\providecommand \bibitemStop [0]{}%
\providecommand \bibitemNoStop [0]{.\EOS\space}%
\providecommand \EOS [0]{\spacefactor3000\relax}%
\providecommand \BibitemShut  [1]{\csname bibitem#1\endcsname}%
\let\auto@bib@innerbib\@empty
%</preamble>
\bibitem [{\citenamefont {Heitzman}(2017)}]{whygooglebest}%
  \BibitemOpen
  \bibfield  {author} {\bibinfo {author} {\bibfnamefont {A.}~\bibnamefont
  {Heitzman}},\ }\bibfield  {title} {\bibinfo {title}
  {\href{https://www.forbes.com/sites/forbesagencycouncil/2017/06/05/how-google-came-to-dominate-search-and-what-the-future-holds/?sh=3a06d8ec3872}{How
  Google Came To Dominate Search And What The Future Holds}},\ }\href@noop {}
  {\bibfield  {journal} {\bibinfo  {journal}
  {https://www.forbes.com/sites/forbesagencycouncil/2017/06/05/how-google-came-to-dominate-search-and-what-the-future-holds/?sh=3a06d8ec3872}\
  } (\bibinfo {year} {2017})}\BibitemShut {NoStop}%
\bibitem [{bes(2020)}]{bestgoogle}%
  \BibitemOpen
  \bibfield  {title} {\bibinfo {title}
  {\href{https://www.towermarketing.net/blog/google-best-search-engine/}{Why
  Google is the Best Search Engine (and Why Businesses Should Care) - In My
  Opinion, by a Tower Alumni}},\ }\href@noop {} {\bibfield  {journal} {\bibinfo
   {journal} {https://www.towermarketing.net/blog/google-best-search-engine/}\
  } (\bibinfo {year} {2020})}\BibitemShut {NoStop}%
\bibitem [{\citenamefont {Rousseau}\ \emph {et~al.}(2008)\citenamefont
  {Rousseau}, \citenamefont {Saint-Aubin}, \citenamefont {Antaya},
  \citenamefont {Ascah-Coallier},\ and\ \citenamefont
  {Hamilton}}]{rousseau2008mathematics}%
  \BibitemOpen
  \bibfield  {author} {\bibinfo {author} {\bibfnamefont {C.}~\bibnamefont
  {Rousseau}}, \bibinfo {author} {\bibfnamefont {Y.}~\bibnamefont
  {Saint-Aubin}}, \bibinfo {author} {\bibfnamefont {H.}~\bibnamefont {Antaya}},
  \bibinfo {author} {\bibfnamefont {I.}~\bibnamefont {Ascah-Coallier}},\ and\
  \bibinfo {author} {\bibfnamefont {C.}~\bibnamefont {Hamilton}},\ }\href@noop
  {} {\emph {\bibinfo {title} {Mathematics and technology}}}\ (\bibinfo
  {publisher} {Springer},\ \bibinfo {year} {2008})\BibitemShut {NoStop}%
\bibitem [{\citenamefont {Rogers}(2002)}]{rogers2002google}%
  \BibitemOpen
  \bibfield  {author} {\bibinfo {author} {\bibfnamefont {I.}~\bibnamefont
  {Rogers}},\ }\bibfield  {title} {\bibinfo {title}
  {\href{chrome-extension://efaidnbmnnnibpcajpcglclefindmkaj/https://cs.wmich.edu/gupta/teaching/cs3310/lectureNotes_cs3310/Pagerank\%20Explained\%20Correctly\%20with\%20Examples_www.cs.princeton.edu_~chazelle_courses_BIB_pagerank.pdf}{The
  Google Pagerank algorithm and how it works}},\ }\href@noop {} {\  (\bibinfo
  {year} {2002})}\BibitemShut {NoStop}%
\bibitem [{\citenamefont
  {https://en.wikipedia.org/wiki/PageRank}()}]{pagerank}%
  \BibitemOpen
  \bibfield  {author} {\bibinfo {author} {\bibnamefont
  {https://en.wikipedia.org/wiki/PageRank}},\ }\bibfield  {title} {\bibinfo
  {title} {\href{https://en.wikipedia.org/wiki/PageRank}{PageRank}},\
  }\href@noop {} {\bibinfo  {journal} {Wikipedia}\ }\BibitemShut {NoStop}%
\bibitem [{\citenamefont {Brin}\ and\ \citenamefont
  {Page}(1998)}]{brin1998anatomy}%
  \BibitemOpen
\bibfield  {journal} {  }\bibfield  {author} {\bibinfo {author} {\bibfnamefont
  {S.}~\bibnamefont {Brin}}\ and\ \bibinfo {author} {\bibfnamefont
  {L.}~\bibnamefont {Page}},\ }\bibfield  {title} {\bibinfo {title} {The
  anatomy of a large-scale hypertextual web search engine},\ }\href@noop {}
  {\bibfield  {journal} {\bibinfo  {journal} {Computer networks and ISDN
  systems}\ }\textbf {\bibinfo {volume} {30}},\ \bibinfo {pages} {107}
  (\bibinfo {year} {1998})}\BibitemShut {NoStop}%
\bibitem [{\citenamefont {Kim}\ and\ \citenamefont
  {Lee}(2002)}]{kim2002improved}%
  \BibitemOpen
  \bibfield  {author} {\bibinfo {author} {\bibfnamefont {S.~J.}\ \bibnamefont
  {Kim}}\ and\ \bibinfo {author} {\bibfnamefont {S.~H.}\ \bibnamefont {Lee}},\
  }\bibfield  {title} {\bibinfo {title} {An improved computation of the
  pagerank algorithm},\ }in\ \href@noop {} {\emph {\bibinfo {booktitle}
  {European Conference on Information Retrieval}}}\ (\bibinfo {organization}
  {Springer},\ \bibinfo {year} {2002})\ pp.\ \bibinfo {pages}
  {73--85}\BibitemShut {NoStop}%
\bibitem [{\citenamefont {Langville}\ and\ \citenamefont
  {Meyer}(2006)}]{langville2006reordering}%
  \BibitemOpen
  \bibfield  {author} {\bibinfo {author} {\bibfnamefont {A.~N.}\ \bibnamefont
  {Langville}}\ and\ \bibinfo {author} {\bibfnamefont {C.~D.}\ \bibnamefont
  {Meyer}},\ }\bibfield  {title} {\bibinfo {title} {A reordering for the
  pagerank problem},\ }\href@noop {} {\bibfield  {journal} {\bibinfo  {journal}
  {SIAM Journal on Scientific Computing}\ }\textbf {\bibinfo {volume} {27}},\
  \bibinfo {pages} {2112} (\bibinfo {year} {2006})}\BibitemShut {NoStop}%
\bibitem [{\citenamefont {Mohan}\ and\ \citenamefont
  {Kurmi}(2017)}]{mohan2017technique}%
  \BibitemOpen
  \bibfield  {author} {\bibinfo {author} {\bibfnamefont {K.}~\bibnamefont
  {Mohan}}\ and\ \bibinfo {author} {\bibfnamefont {J.}~\bibnamefont {Kurmi}},\
  }\bibfield  {title} {\bibinfo {title} {A technique to improved page rank
  algorithm in perspective to optimized normalization technique.},\ }\href@noop
  {} {\bibfield  {journal} {\bibinfo  {journal} {International Journal of
  Advanced Research in Computer Science}\ }\textbf {\bibinfo {volume} {8}}
  (\bibinfo {year} {2017})}\BibitemShut {NoStop}%
\bibitem [{\citenamefont {S{\'a}nchez-Burillo}\ \emph
  {et~al.}(2012)\citenamefont {S{\'a}nchez-Burillo}, \citenamefont {Duch},
  \citenamefont {G{\'o}mez-Gardenes},\ and\ \citenamefont
  {Zueco}}]{sanchez2012quantum}%
  \BibitemOpen
  \bibfield  {author} {\bibinfo {author} {\bibfnamefont {E.}~\bibnamefont
  {S{\'a}nchez-Burillo}}, \bibinfo {author} {\bibfnamefont {J.}~\bibnamefont
  {Duch}}, \bibinfo {author} {\bibfnamefont {J.}~\bibnamefont
  {G{\'o}mez-Gardenes}},\ and\ \bibinfo {author} {\bibfnamefont
  {D.}~\bibnamefont {Zueco}},\ }\bibfield  {title} {\bibinfo {title} {Quantum
  navigation and ranking in complex networks},\ }\href@noop {} {\bibfield
  {journal} {\bibinfo  {journal} {Scientific Reports}\ }\textbf {\bibinfo
  {volume} {2}},\ \bibinfo {pages} {1} (\bibinfo {year} {2012})}\BibitemShut
  {NoStop}%
\bibitem [{\citenamefont {Whitfield}\ \emph {et~al.}(2010)\citenamefont
  {Whitfield}, \citenamefont {Rodr\'{\i}guez-Rosario},\ and\ \citenamefont
  {Aspuru-Guzik}}]{QSW_def}%
  \BibitemOpen
  \bibfield  {author} {\bibinfo {author} {\bibfnamefont {J.~D.}\ \bibnamefont
  {Whitfield}}, \bibinfo {author} {\bibfnamefont {C.~A.}\ \bibnamefont
  {Rodr\'{\i}guez-Rosario}},\ and\ \bibinfo {author} {\bibfnamefont
  {A.}~\bibnamefont {Aspuru-Guzik}},\ }\bibfield  {title} {\bibinfo {title}
  {\href{https://link.aps.org/doi/10.1103/PhysRevA.81.022323}{Quantum
  stochastic walks: A generalization of classical random walks and quantum
  walks}},\ }\href@noop {} {\bibfield  {journal} {\bibinfo  {journal} {Physical
  Review A}\ }\textbf {\bibinfo {volume} {81}},\ \bibinfo {pages} {022323}
  (\bibinfo {year} {2010})}\BibitemShut {NoStop}%
\bibitem [{\citenamefont {Portugal}(2018)}]{Portugal}%
  \BibitemOpen
  \bibfield  {author} {\bibinfo {author} {\bibfnamefont {R.}~\bibnamefont
  {Portugal}},\ }\href {https://doi.org/10.1007/978-3-319-97813-0_3} {\emph
  {\bibinfo {title} {Quantum Walks and Search Algorithms}}}\ (\bibinfo
  {publisher} {Springer International Publishing},\ \bibinfo {address} {Cham},\
  \bibinfo {year} {2018})\BibitemShut {NoStop}%
\bibitem [{\citenamefont {Aharonov}\ \emph {et~al.}(1993)\citenamefont
  {Aharonov}, \citenamefont {Davidovich},\ and\ \citenamefont
  {Zagury}}]{coineddef}%
  \BibitemOpen
  \bibfield  {author} {\bibinfo {author} {\bibfnamefont {Y.}~\bibnamefont
  {Aharonov}}, \bibinfo {author} {\bibfnamefont {L.}~\bibnamefont
  {Davidovich}},\ and\ \bibinfo {author} {\bibfnamefont {N.}~\bibnamefont
  {Zagury}},\ }\bibfield  {title} {\bibinfo {title}
  {\href{https://link.aps.org/doi/10.1103/PhysRevA.48.1687}{Quantum random
  walks}},\ }\href@noop {} {\bibfield  {journal} {\bibinfo  {journal} {Physical
  Review A}\ }\textbf {\bibinfo {volume} {48}},\ \bibinfo {pages} {1687}
  (\bibinfo {year} {1993})}\BibitemShut {NoStop}%
\bibitem [{\citenamefont {Farhi}\ and\ \citenamefont {Gutmann}(1998)}]{Farhi}%
  \BibitemOpen
  \bibfield  {author} {\bibinfo {author} {\bibfnamefont {E.}~\bibnamefont
  {Farhi}}\ and\ \bibinfo {author} {\bibfnamefont {S.}~\bibnamefont
  {Gutmann}},\ }\bibfield  {title} {\bibinfo {title}
  {\href{https://link.aps.org/doi/10.1103/PhysRevA.58.915}{Quantum computation
  and decision trees}},\ }\href@noop {} {\bibfield  {journal} {\bibinfo
  {journal} {Physical Review A}\ }\textbf {\bibinfo {volume} {58}},\ \bibinfo
  {pages} {915} (\bibinfo {year} {1998})}\BibitemShut {NoStop}%
\bibitem [{\citenamefont {Loke}\ \emph {et~al.}(2017)\citenamefont {Loke},
  \citenamefont {Tang}, \citenamefont {Rodriguez}, \citenamefont {Small},\ and\
  \citenamefont {Wang}}]{loke2017comparing}%
  \BibitemOpen
  \bibfield  {author} {\bibinfo {author} {\bibfnamefont {T.}~\bibnamefont
  {Loke}}, \bibinfo {author} {\bibfnamefont {J.~W.}\ \bibnamefont {Tang}},
  \bibinfo {author} {\bibfnamefont {J.}~\bibnamefont {Rodriguez}}, \bibinfo
  {author} {\bibfnamefont {M.}~\bibnamefont {Small}},\ and\ \bibinfo {author}
  {\bibfnamefont {J.~B.}\ \bibnamefont {Wang}},\ }\bibfield  {title} {\bibinfo
  {title} {Comparing classical and quantum pageranks},\ }\href@noop {}
  {\bibfield  {journal} {\bibinfo  {journal} {Quantum Information Processing}\
  }\textbf {\bibinfo {volume} {16}},\ \bibinfo {pages} {1} (\bibinfo {year}
  {2017})}\BibitemShut {NoStop}%
\bibitem [{\citenamefont {Falloon}\ \emph
  {et~al.}(2017{\natexlab{a}})\citenamefont {Falloon}, \citenamefont
  {Rodriguez},\ and\ \citenamefont {Wang}}]{QSWalk}%
  \BibitemOpen
  \bibfield  {author} {\bibinfo {author} {\bibfnamefont {P.~E.}\ \bibnamefont
  {Falloon}}, \bibinfo {author} {\bibfnamefont {J.}~\bibnamefont {Rodriguez}},\
  and\ \bibinfo {author} {\bibfnamefont {J.~B.}\ \bibnamefont {Wang}},\
  }\bibfield  {title} {\bibinfo {title} {Qswalk: A mathematica package for
  quantum stochastic walks on arbitrary graphs},\ }\href
  {https://doi.org/https://doi.org/10.1016/j.cpc.2017.03.014} {\bibfield
  {journal} {\bibinfo  {journal} {Computer Physics Communications}\ }\textbf
  {\bibinfo {volume} {217}},\ \bibinfo {pages} {162 } (\bibinfo {year}
  {2017}{\natexlab{a}})}\BibitemShut {NoStop}%
\bibitem [{\citenamefont {Kossakowski}(1972)}]{KOSSAKOWSKI}%
  \BibitemOpen
  \bibfield  {author} {\bibinfo {author} {\bibfnamefont {A.}~\bibnamefont
  {Kossakowski}},\ }\bibfield  {title} {\bibinfo {title}
  {\href{http://www.sciencedirect.com/science/article/pii/0034487772900109}{On
  quantum statistical mechanics of non-Hamiltonian systems}},\ }\href@noop {}
  {\bibfield  {journal} {\bibinfo  {journal} {Reports on Mathematical Physics}\
  }\textbf {\bibinfo {volume} {3}},\ \bibinfo {pages} {247 } (\bibinfo {year}
  {1972})}\BibitemShut {NoStop}%
\bibitem [{\citenamefont {Lindblad}(1976)}]{Lindblad}%
  \BibitemOpen
  \bibfield  {author} {\bibinfo {author} {\bibfnamefont {G.}~\bibnamefont
  {Lindblad}},\ }\bibfield  {title} {\bibinfo {title}
  {\href{https://doi.org/10.1007/BF01608499}{On the generators of quantum
  dynamical semigroups}},\ }\href@noop {} {\bibfield  {journal} {\bibinfo
  {journal} {Communications in Mathematical Physics}\ }\textbf {\bibinfo
  {volume} {48}},\ \bibinfo {pages} {119} (\bibinfo {year} {1976})}\BibitemShut
  {NoStop}%
\bibitem [{\citenamefont {Kendon}(2007)}]{kendon}%
  \BibitemOpen
  \bibfield  {author} {\bibinfo {author} {\bibfnamefont {V.}~\bibnamefont
  {Kendon}},\ }\bibfield  {title} {\bibinfo {title} {Decoherence in quantum
  walks - a review},\ }\href {https://doi.org/10.1017/S0960129507006354}
  {\bibfield  {journal} {\bibinfo  {journal} {Mathematical Structures in
  Computer Science}\ }\textbf {\bibinfo {volume} {17}},\ \bibinfo {pages}
  {1169} (\bibinfo {year} {2007})}\BibitemShut {NoStop}%
\bibitem [{\citenamefont {Dudhe}\ \emph {et~al.}(2021)\citenamefont {Dudhe},
  \citenamefont {Sahoo},\ and\ \citenamefont {Benjamin}}]{testingspeedup}%
  \BibitemOpen
  \bibfield  {author} {\bibinfo {author} {\bibfnamefont {N.}~\bibnamefont
  {Dudhe}}, \bibinfo {author} {\bibfnamefont {P.~K.}\ \bibnamefont {Sahoo}},\
  and\ \bibinfo {author} {\bibfnamefont {C.}~\bibnamefont {Benjamin}},\
  }\bibfield  {title} {\bibinfo {title} {Testing quantum speedups in exciton
  transport through a photosynthetic complex using quantum stochastic walks},\
  }\href {https://doi.org/10.1039/D1CP02727A} {\bibfield  {journal} {\bibinfo
  {journal} {Phys. Chem. Chem. Phys.}\ }\textbf {\bibinfo {volume} {24}},\
  \bibinfo {pages} {2601} (\bibinfo {year} {2021})}\BibitemShut {NoStop}%
\bibitem [{\citenamefont {Watts}\ and\ \citenamefont
  {Strogatz}(1998)}]{watts1998collective}%
  \BibitemOpen
  \bibfield  {author} {\bibinfo {author} {\bibfnamefont {D.~J.}\ \bibnamefont
  {Watts}}\ and\ \bibinfo {author} {\bibfnamefont {S.~H.}\ \bibnamefont
  {Strogatz}},\ }\bibfield  {title} {\bibinfo {title} {Collective dynamics of
  `small-world'networks},\ }\href@noop {} {\bibfield  {journal} {\bibinfo
  {journal} {Nature}\ }\textbf {\bibinfo {volume} {393}},\ \bibinfo {pages}
  {440} (\bibinfo {year} {1998})}\BibitemShut {NoStop}%
\bibitem [{\citenamefont {Chen}(2018)}]{chen2018models}%
  \BibitemOpen
  \bibfield  {author} {\bibinfo {author} {\bibfnamefont {Y.}~\bibnamefont
  {Chen}},\ }\href@noop {} {\bibinfo {title} {Models of complex networks}}
  (\bibinfo {year} {2018})\BibitemShut {NoStop}%
\bibitem [{\citenamefont {Zachary}(1977)}]{zachary1977information}%
  \BibitemOpen
  \bibfield  {author} {\bibinfo {author} {\bibfnamefont {W.~W.}\ \bibnamefont
  {Zachary}},\ }\bibfield  {title} {\bibinfo {title} {An information flow model
  for conflict and fission in small groups},\ }\href@noop {} {\bibfield
  {journal} {\bibinfo  {journal} {Journal of Anthropological Research}\
  }\textbf {\bibinfo {volume} {33}},\ \bibinfo {pages} {452} (\bibinfo {year}
  {1977})}\BibitemShut {NoStop}%
\bibitem [{\citenamefont {Weisstein}(2002)}]{weisstein2002bernoulli}%
  \BibitemOpen
  \bibfield  {author} {\bibinfo {author} {\bibfnamefont {E.~W.}\ \bibnamefont
  {Weisstein}},\ }\bibfield  {title} {\bibinfo {title} {Bernoulli
  distribution},\ }\href@noop {} {\bibfield  {journal} {\bibinfo  {journal}
  {https://mathworld. wolfram. com/}\ } (\bibinfo {year} {2002})}\BibitemShut
  {NoStop}%
\bibitem [{\citenamefont {Albert}\ and\ \citenamefont
  {Barab{\'a}si}(2002)}]{albert2002statistical}%
  \BibitemOpen
  \bibfield  {author} {\bibinfo {author} {\bibfnamefont {R.}~\bibnamefont
  {Albert}}\ and\ \bibinfo {author} {\bibfnamefont {A.-L.}\ \bibnamefont
  {Barab{\'a}si}},\ }\bibfield  {title} {\bibinfo {title} {Statistical
  mechanics of complex networks},\ }\href@noop {} {\bibfield  {journal}
  {\bibinfo  {journal} {Reviews of Modern Physics}\ }\textbf {\bibinfo {volume}
  {74}},\ \bibinfo {pages} {47} (\bibinfo {year} {2002})}\BibitemShut {NoStop}%
\bibitem [{\citenamefont {Price}(1965)}]{price1965networks}%
  \BibitemOpen
  \bibfield  {author} {\bibinfo {author} {\bibfnamefont {D.~J. D.~S.}\
  \bibnamefont {Price}},\ }\bibfield  {title} {\bibinfo {title} {Networks of
  scientific papers: The pattern of bibliographic references indicates the
  nature of the scientific research front},\ }\href@noop {} {\bibfield
  {journal} {\bibinfo  {journal} {Science}\ }\textbf {\bibinfo {volume}
  {149}},\ \bibinfo {pages} {510} (\bibinfo {year} {1965})}\BibitemShut
  {NoStop}%
\bibitem [{\citenamefont {Price}(1976)}]{price1976general}%
  \BibitemOpen
  \bibfield  {author} {\bibinfo {author} {\bibfnamefont {D.~d.~S.}\
  \bibnamefont {Price}},\ }\bibfield  {title} {\bibinfo {title} {A general
  theory of bibliometric and other cumulative advantage processes},\
  }\href@noop {} {\bibfield  {journal} {\bibinfo  {journal} {Journal of the
  American Society for Information Science}\ }\textbf {\bibinfo {volume}
  {27}},\ \bibinfo {pages} {292} (\bibinfo {year} {1976})}\BibitemShut
  {NoStop}%
\bibitem [{\citenamefont {Barth{\'e}lemy}(2011)}]{barthelemy2011spatial}%
  \BibitemOpen
  \bibfield  {author} {\bibinfo {author} {\bibfnamefont {M.}~\bibnamefont
  {Barth{\'e}lemy}},\ }\bibfield  {title} {\bibinfo {title} {Spatial
  networks},\ }\href@noop {} {\bibfield  {journal} {\bibinfo  {journal}
  {Physics Reports}\ }\textbf {\bibinfo {volume} {499}},\ \bibinfo {pages} {1}
  (\bibinfo {year} {2011})}\BibitemShut {NoStop}%
\bibitem [{\citenamefont {Amaral}\ \emph {et~al.}(2000)\citenamefont {Amaral},
  \citenamefont {Scala}, \citenamefont {Barthelemy},\ and\ \citenamefont
  {Stanley}}]{amaral2000classes}%
  \BibitemOpen
  \bibfield  {author} {\bibinfo {author} {\bibfnamefont {L.~A.~N.}\
  \bibnamefont {Amaral}}, \bibinfo {author} {\bibfnamefont {A.}~\bibnamefont
  {Scala}}, \bibinfo {author} {\bibfnamefont {M.}~\bibnamefont {Barthelemy}},\
  and\ \bibinfo {author} {\bibfnamefont {H.~E.}\ \bibnamefont {Stanley}},\
  }\bibfield  {title} {\bibinfo {title} {Classes of small-world networks},\
  }\href@noop {} {\bibfield  {journal} {\bibinfo  {journal} {Proceedings of the
  National Academy of Sciences}\ }\textbf {\bibinfo {volume} {97}},\ \bibinfo
  {pages} {11149} (\bibinfo {year} {2000})}\BibitemShut {NoStop}%
\bibitem [{\citenamefont {Chehreghani}\ and\ \citenamefont
  {Chehreghani}(2014)}]{chehreghani2014modeling}%
  \BibitemOpen
  \bibfield  {author} {\bibinfo {author} {\bibfnamefont {M.~H.}\ \bibnamefont
  {Chehreghani}}\ and\ \bibinfo {author} {\bibfnamefont {M.~H.}\ \bibnamefont
  {Chehreghani}},\ }\bibfield  {title} {\bibinfo {title} {Modeling transitivity
  in complex networks},\ }\href@noop {} {\bibfield  {journal} {\bibinfo
  {journal} {arXiv:1411.0958}\ } (\bibinfo {year} {2014})}\BibitemShut
  {NoStop}%
\bibitem [{zac()}]{zacharynetwork}%
  \BibitemOpen
  \bibfield  {title} {\bibinfo {title}
  {\href{https://reference.wolfram.com/language/example/AnalyzeSocialNetworks.html}{Analyze
  Social Networks}},\ }\href@noop {} {\bibinfo  {journal}
  {https://reference.wolfram.com/language/example/AnalyzeSocialNetworks.html}\
  }\BibitemShut {NoStop}%
\bibitem [{ber()}]{bernoulligraph}%
  \BibitemOpen
\bibfield  {journal} {  }\bibfield  {title} {\bibinfo {title}
  {\href{https://reference.wolfram.com/language/ref/BernoulliGraphDistribution.html}{Bernoulli
  Graph Distribution}},\ }\href@noop {} {\bibinfo  {journal}
  {https://reference.wolfram.com/language/ref/BernoulliGraphDistribution.html}\
  }\BibitemShut {NoStop}%
\bibitem [{\citenamefont {Song}\ and\ \citenamefont
  {Wang}(2014)}]{song2014simple}%
  \BibitemOpen
\bibfield  {journal} {  }\bibfield  {author} {\bibinfo {author} {\bibfnamefont
  {H.~F.}\ \bibnamefont {Song}}\ and\ \bibinfo {author} {\bibfnamefont {X.-J.}\
  \bibnamefont {Wang}},\ }\bibfield  {title} {\bibinfo {title} {Simple,
  distance-dependent formulation of the watts-strogatz model for directed and
  undirected small-world networks},\ }\href@noop {} {\bibfield  {journal}
  {\bibinfo  {journal} {Physical Review E}\ }\textbf {\bibinfo {volume} {90}},\
  \bibinfo {pages} {062801} (\bibinfo {year} {2014})}\BibitemShut {NoStop}%
\bibitem [{\citenamefont {Kimble}(2008)}]{kimble2008quantum}%
  \BibitemOpen
  \bibfield  {author} {\bibinfo {author} {\bibfnamefont {H.~J.}\ \bibnamefont
  {Kimble}},\ }\bibfield  {title} {\bibinfo {title} {The quantum internet},\
  }\href@noop {} {\bibfield  {journal} {\bibinfo  {journal} {Nature}\ }\textbf
  {\bibinfo {volume} {453}},\ \bibinfo {pages} {1023} (\bibinfo {year}
  {2008})}\BibitemShut {NoStop}%
\bibitem [{\citenamefont {Schuld}\ \emph {et~al.}(2014)\citenamefont {Schuld},
  \citenamefont {Sinayskiy},\ and\ \citenamefont
  {Petruccione}}]{schuld2014quantum}%
  \BibitemOpen
  \bibfield  {author} {\bibinfo {author} {\bibfnamefont {M.}~\bibnamefont
  {Schuld}}, \bibinfo {author} {\bibfnamefont {I.}~\bibnamefont {Sinayskiy}},\
  and\ \bibinfo {author} {\bibfnamefont {F.}~\bibnamefont {Petruccione}},\
  }\bibfield  {title} {\bibinfo {title} {Quantum walks on graphs representing
  the firing patterns of a quantum neural network},\ }\href@noop {} {\bibfield
  {journal} {\bibinfo  {journal} {Physical Review A}\ }\textbf {\bibinfo
  {volume} {89}},\ \bibinfo {pages} {032333} (\bibinfo {year}
  {2014})}\BibitemShut {NoStop}%
\bibitem [{\citenamefont {Falloon}\ \emph
  {et~al.}(2017{\natexlab{b}})\citenamefont {Falloon}, \citenamefont
  {Rodriguez},\ and\ \citenamefont {Wang}}]{QSWalkdownload}%
  \BibitemOpen
  \bibfield  {author} {\bibinfo {author} {\bibfnamefont {P.~E.}\ \bibnamefont
  {Falloon}}, \bibinfo {author} {\bibfnamefont {J.}~\bibnamefont {Rodriguez}},\
  and\ \bibinfo {author} {\bibfnamefont {J.~B.}\ \bibnamefont {Wang}},\
  }\href@noop {} {\emph {\bibinfo {title}
  {\href{https://data.mendeley.com/datasets/8rwd3j9zhk/1}{QSWalkdownload}}}}
  (\bibinfo {year} {2017}{\natexlab{b}})\BibitemShut {NoStop}%
\end{thebibliography}%
\end{document}